\documentclass[prd,reprint,superscriptaddress,nofootinbib]{revtex4}

\usepackage{epsfig,subfigure}
\usepackage{epstopdf}
\usepackage{graphicx}
\usepackage{amssymb}
\usepackage{amsxtra}
\usepackage{amsmath}
\usepackage{color}
\usepackage{hyperref}
\usepackage{algpseudocode}
\usepackage{soul}
\usepackage{tikz-feynman}

\newcommand{\bea}{\begin{eqnarray}}
\newcommand{\eea}{\end{eqnarray}}

\def\beq{\begin{equation}}
\def\eeq{\end{equation}}

\newcommand{\lsim}{\lesssim}

\begin{document}

\title{Multi-photon decays of the Higgs boson at the LHC}

\author{Samuel D. Lane}
\email{samuellane@kaist.ac.kr}
\affiliation{Department of Physics, KAIST, Daejeon 34141, Korea}

\author{Hye-Sung Lee}
\email{hyesung.lee@kaist.ac.kr}
\affiliation{Department of Physics, KAIST, Daejeon 34141, Korea}

\author{Ian M. Lewis}
\email{ian.lewis@ku.edu}
\affiliation{Department of Physics and Astronomy, University of Kansas, Lawrence, Kansas 66045, USA }
\affiliation{Department of Physics and Astronomy, University of Pittsburgh, Pittsburgh, Pennsylvania 15260, USA}


\begin{abstract}
Many new physics scenarios predict multi-photon Higgs resonances. One such scenario is the dark axion portal model. The primary decay chain  that we study is the Higgs to dark photon ($\gamma_D$) pairs that subsequently decay into a photon ($\gamma$) and an axion-like particle ($a$). The axion-like particles then decay into photon pairs. Hence, the signal is a six-photon Higgs decay: $h\rightarrow \gamma_D\,\gamma_D\rightarrow 2\,\gamma 2\,a\rightarrow 6\gamma$. However, depending on the relevant kinematics, the photons can become well-collimated and appear as photon-jets (multiple photons that appear as a single photon in the detector) or $\xi$-jets (non-isolated multi-photon signals that do not pass the isolation criterion). These effects cause the true six-photon resonance to appear as other multi-photon signals, such as two and four photons.   We classify the mass regions where two, four, and six-photon resonances dominate. The four-photon signal is particularly interesting.  These events mainly occur when the photons from the axion-like particles are collimated into photon-jets.  The decay of the dark photon is then $\gamma_D\rightarrow \gamma a\rightarrow \gamma+\gamma$-{\rm jet}, which is an apparent violation of the Landau-Yang theorem.  We show that current measurements of $h\rightarrow 2\gamma$ and searches for $h\rightarrow 4\gamma$ at the Large Hadron Collider (LHC) can limit ${\rm BR}(h\rightarrow \gamma_D\gamma_D)\lesssim 10^{-3}$.  This model also motivates new searches for Higgs decays into six isolated photons or $\xi$-jets at the LHC.  While there are currently no dedicated searches, we show that many of the Higgs to six isolated photons or $\xi$-jet events could pass two or three-photon triggers.  That is, new physics could be found by reanalyzing existing data.  These multi-photon signals provide excellent footing to explore new physics at the LHC and beyond.
\end{abstract}

\maketitle

\section{Introduction}
The Standard Model (SM) of particle physics has had great success in predicting experimental signatures of the known fundamental particles. Yet, there are observed phenomena that remain unexplained by the SM, such as dark matter.  Many solutions to the dark matter problem postulate that there is a dark sector that couples to the SM through a ``portal''. 
These portals include the vector portal, neutrino portal, Higgs portal, axion portal, and dark axion portal. (For a review on some of these portals, see Ref.~\cite{Essig:2013lka}.) There have been vast efforts to look for evidence of these new states and couplings across many experiments. Some of the most recent results pertaining to beyond the standard model (BSM) searches and strategies have been summarized in the Snowmass reports \cite{Narain:2022qud,
Bose:2022obr}. The reports also consider how the new portal states affect precision Higgs and electroweak  (EW) measurements \cite{Belloni:2022due, Dawson:2022zbb}.  
  
Many of these portals and dark sector models involve introducing light particles \cite{Graham:2010ca, Curtin:2014cca, DAgnolo:2015ujb, Liu:2016qwd, Dror:2017ehi, Green:2017ybv, Dvorkin:2019zdi, Berlin:2018bsc}, which will often couple to the SM Higgs boson \cite{ Davoudiasl:2012ig, Davoudiasl:2013aya, Davoudiasl:2015bua, Lu:2017uur, Muhlleitner:2017dkd, Ilnicka:2018def, Rizzo:2018vlb}. Some of these new states might also decay primarily to photons, depending on the coupling and mass structure. In this scenario, one would generically expect multi-photon decays of the Higgs boson mediated by light resonances. These multi-photon resonances have the potential to provide relatively clean signals of new physics at hadron colliders. Indeed, it was the diphoton decay that provided some of the first evidence of the Higgs boson at the Large Hadron Collider (LHC) \cite{CMS:2012qbp, ATLAS:2012yve}. As a result of their clean signature, there have been many searches at the LHC for two and four-photon resonances 
\cite{ATLAS:2018dfo, ATLAS:2022tnm,ATLAS:2022abz, CMS:2022fyt, CMS:2022xxa, Knapen:2021elo, Cepeda:2021rql, CMS:2021kom, ATLAS:2021mbt, Stefaniak:2019hvg, Chakraborty:2017mbz, Sheff:2020jyw, Cacciapaglia:2022bax}.
However, there have yet to be any significant signals outside the SM expectations \cite{CMS:2022wpo, ATLAS:2022vkf, ATLAS:2023tkt, Salam:2022izo}.

Light states originating from Higgs decays can provide many challenges.  Such particles are naturally boosted and could produce well-collimated photons. As has been noted by many authors in the literature, when the photons become highly collimated, they can produce so-called photon-jets ($\gamma$-jets)~\cite{Draper:2012xt, Ellis:2012zp}, where the photons are too collimated to be distinguished in a detector. Furthermore, there has been recent interest in intermediately separated photons ($\xi$-jets) \cite{ATLAS:2018dfo, Chakraborty:2017mbz, Sheff:2020jyw}. These photons are too collimated to pass the isolation requirements of the experiments but separated enough such that they cannot be mistaken for a single photon~ \cite{CMS:2009nxa, ATLAS:2019qmc}. As such, these $\xi$-jets provide particular difficulty in passing trigger requirements and reconstructing the Higgs mass. 

\begin{figure}
\begin{center}
\subfigure[]{\label{fig:photonX}\includegraphics[width=0.4\textwidth,clip]{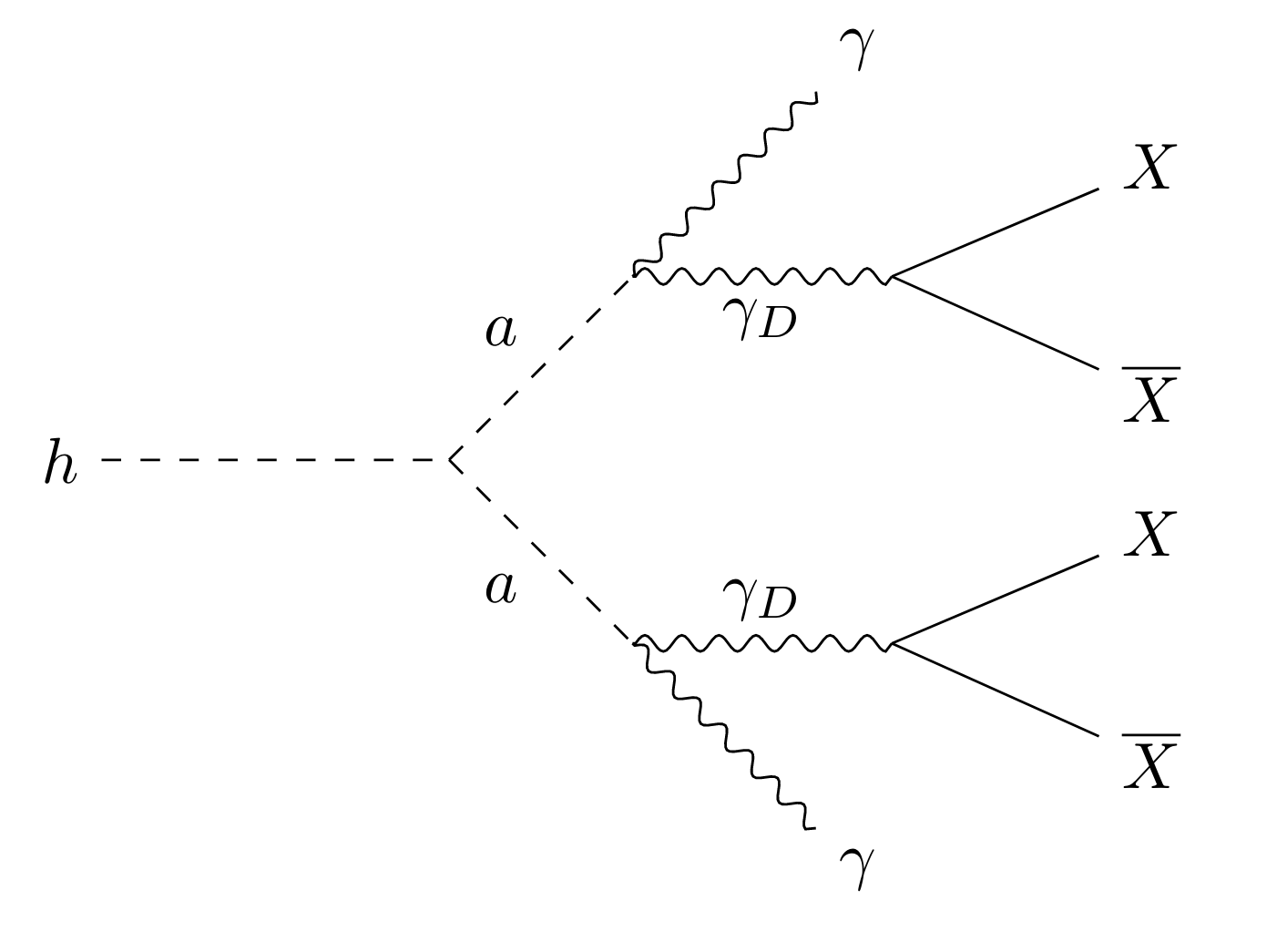}}
\subfigure[]{\label{fig:sixphoton}\includegraphics[width=0.4\textwidth, clip]{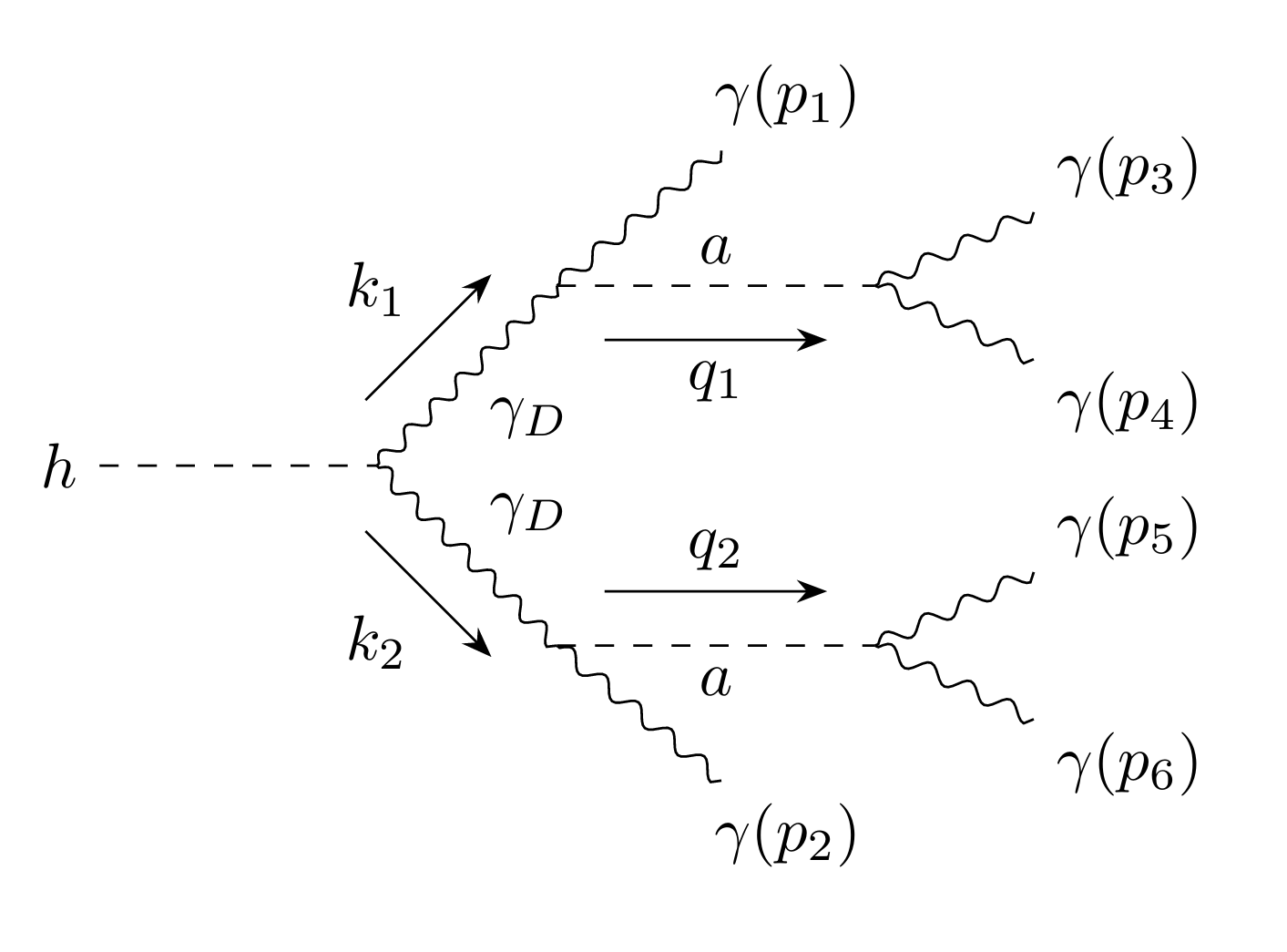}}
\end{center}
\caption{\label{fig:HiggsDecay} Tree level Feynman diagrams in the dark axion portal model for Higgs decays (a) to photons and other particles and (b) to six-photons. }
\end{figure}

In this paper, we study multi-photon Higgs decays in the dark axion portal model~\cite{Kaneta:2016wvf}. This model introduces an axion-like particle (ALP) as well as a dark photon, and their connection through anomaly triangles. Provided the ALP and dark photon are sufficiently light, they can be produced via decays of the observed Higgs boson $h$.  If the ALP, $a$, is more massive than the dark photon, $\gamma_D$, these decays can be of the form $h\rightarrow aa \rightarrow (\gamma \gamma_D)(\gamma \gamma_D)\rightarrow 2\gamma+X$ where X are some kinematically accessible SM or dark sector particles. The particles $X$ need not be all the same species. Fig.~\ref{fig:photonX} shows a representative diagram of this decay.

Particles that induce a new loop coupling $a-\gamma-\gamma_D$ will also induce a $a-\gamma-\gamma$ coupling.  Hence, if the ALP is less massive than the dark photon, the axion portal model will provide the multi-photon Higgs decay 
\begin{eqnarray}
h \rightarrow \gamma_D \gamma_D \rightarrow (\gamma a) (\gamma a) \rightarrow 6 \gamma,
\end{eqnarray}
as shown in Fig.~\ref{fig:sixphoton}. In this paper, we investigate precisely this signal.  We classify the possible signatures, recast current LHC results as constraints on this model, and propose new signatures to search for at the LHC.

Depending on the masses of both the ALP and dark photon, the final state photons can receive a significant boost.  Overall, the main factor in determining the different observed signals is the kinematics of the intermediate particles and not the values of the couplings. These six-photon signals can manifest as a combination of photons, photon-jets, and $\xi$-jets. We will see the most relevant signals at the LHC are the Higgs apparent decays into six, four, and two-photons, where the four and two-photon signatures often occur via $\gamma$-jets. Many of these signals also involve $\xi$-jets: zero-photon plus two $\xi$-jets, two-photon plus two $\xi$-jets, three-photon plus one $\xi$-jet, or four-photon plus one $\xi$-jet. These signals will not pass the standard analysis, where the $\xi$-jets are not identified as photons. 

Of particular interest is the case where the mass of the ALP is much less than the mass of the dark photon, but the dark photon is not too boosted: $m_a\ll m_{\gamma_D}\sim 10-60$~GeV.  This signal has two photon-jets and two photons.  The $\gamma$-jets originate from the ALP decay. Hence, the dark photon decays, $\gamma_D \rightarrow \gamma a \rightarrow \gamma+\gamma$-jet, are apparently into two photons. This would appear to be a violation of the Landau-Yang theorem, which states that a massive vector particle cannot decay into two massless vector particles \cite{Landau:1948kw,Yang:1950rg}.  While this is not a true violation of the Landau-Yang theorem, it is indistinguishable from one at the detector level. 
Other examples of the apparent Landau-Yang violation can be found in Refs.~\cite{Toro:2012sv,Chala:2015cev}.

The six-isolated photon decay proves particularly challenging to detect. We will make suggestions for how to improve the detector efficiencies of these signals. Furthermore, some interesting signals where the dominant branching fraction includes $\xi$-jets are discussed.  Such signals are not currently searched for at the LHC.  As we will show, many of the six-isolated photon and $\xi$-jet events could have passed the two or three-photon LHC triggers.  That is, these signals could be discovered by a reanalysis of existing data.

The model details are provided in Sec.~\ref{sec:theory}. The signal categories, descriptions of multi-photon objects, and jet reconstruction algorithms are provided in Sec.~\ref{sec:sig}. The various signals and search strategies along with the detector response are discussed in Sec.~\ref{sec:result}. There, we also provide constraints on branching fractions from recast experimental searches. 
Finally, a brief summary and outlook are given in Sec.~\ref{sec:conc}. Appendix~\ref{sec:BR} contains some formulas for the calculation of the branching ratios.

\section{Model Description}
\label{sec:theory}\subsection{Model}
There are many models that have new light particles that primarily decay to photons but also couple to the Higgs. Such models produce multi-photon resonances \cite{Draper:2012xt, Curtin:2013fra, Davoudiasl:2013aya,Ferretti:2013kya, Dimopoulos:2016lvn, Ferretti:2016upr, Bellazzini:2017neg, Gaillard:2018xgk}. The dark KSVZ model, which realized the dark axion portal, is a well-motivated example \cite{Kaneta:2016wvf}.
Although the original focus of the KSVZ model was on the QCD axion, it can be extended to the ALP case straightforwardly. Further, it has all the ingredients we need for our analysis: dark Higgs, dark photon, and ALP.
For simplicity, we will use the phenomenological model described below, regardless of the underlying UV theory.
The masses and couplings of the new particles are considered independent of each other, and we do not address the strong $CP$ problem.

Starting with the SM Lagrangian, two fields are added: an ALP and a massive dark photon. The kinetic and mass terms for the ALP and the dark photon are
\begin{align}
\mathcal{L}_{\text{kinetic+mass}} = -\frac{1}{4} F_D^{\mu\nu} F_{D\mu \nu} + \frac{1}{2}m_{\gamma_D}^2 \gamma_D^{\mu} \gamma_{D\mu} + \frac{1}{2} \partial_{\mu} a \partial^{\mu} a - \frac{1}{2}m_a^2 a^2,
\end{align}
where $F_D^{\mu\nu}$ is the field strength tensor for the dark photon, $m_{\gamma_D}$ is the mass of the dark photon, and $m_a$ the mass of the ALP.  The relevant portal couplings among the SM fields, the ALP, and the dark photon are parameterized by
\begin{align}
\mathcal{L}_{\text{portal}} =  
\frac{G_{a \gamma \gamma}}{4} a F^{\mu \nu} \tilde{F}_{\mu\nu} 
+ \frac{G_{a \gamma \gamma_D}}{2} a F^{\mu \nu} \tilde{F}_{D \mu\nu} 
+ \frac{G_{a \gamma_D \gamma_D}}{4} a F_D^{\mu \nu} \tilde{F}_{D \mu\nu}
+ \frac{\lambda_{h\gamma_D\gamma_D}}{2} h \gamma_D^{\mu} \gamma_{D\mu},
\end{align}
where $F^{\mu\nu}$ is the field strength tensor for the SM photon, and the tilde notation indicates the dual to the field strength tensors.
$G_{a \gamma \gamma}$ is the axion portal, $G_{a \gamma \gamma_D}$ is the dark axion portal, and $\lambda_{h\gamma_D\gamma_D}$ originates from the Higgs portal (mixing of the SM Higgs doublet and the dark Higgs).
For simplicity, we do not consider the ALP couplings to other SM particles such as gluons, which can be avoided by assuming the fermions in the anomaly triangles are singlets under $SU(3)_C$.  
The full Lagrangian is given by
\begin{align}
\mathcal{L}_{\text{total}} = 
\mathcal{L}_{\text{SM}} 
+ \mathcal{L}_{\text{kinetic+mass} } 
+  \mathcal{L}_{\text{portal} },
\end{align}
where $\mathcal{L}_{\text{SM}}$ is the SM Lagrangian.

We will assume the dark photon receives its mass through a dark Higgs mechanism with dark vacuum expectation value $v_D$. Its mass will be related to the dark gauge coupling $g_D$ via $m_{\gamma_D} \sim g_D v_D$. The SM and dark Higgs bosons will mix via a mixing angle $\theta$. Hence, the dark Higgs will couple to the SM fermions and gauge bosons like the SM Higgs, but suppressed by a factor of $\sin\theta$.  It is then possible to produce the dark Higgs at colliders.  
Similarly, the scalar mixing will universally suppress the observed Higgs couplings to the SM fermions and gauge bosons.  Hence, the mixing angle is constrained by the Higgs and precision EW measurements, and scalar searches at the LHC and LEP\cite{Bowen:2007ia, Barger:2007im, Bechtle:2013xfa, Robens:2015gla, Lopez-Val:2014jva, Bechtle:2020uwn}. We will consider the implications of these constraints below. 

One can also see how the scalar mixing induces the Higgs-dark photon couplings.  Generically, the dark Higgs couples to a pair of dark photons as $m_{\gamma_D}^2/v_D$.  Hence, the scalar mixing angle will also induce a coupling between the observed Higgs to the dark photon via the scalar mixing:
\begin{align}
\lambda_{h\gamma_D\gamma_D}  \sim \sin \theta \frac{m_{\gamma_D}^2}{v_D}.\label{eq:HiggsDarkPhoton}
\end{align}

Finally, we assume the kinetic mixing between the photon and the dark photon is zero.
The fermions inside the anomaly triangle that produce the dark axion portal ($G_{a\gamma\gamma_D}$) can also induce the kinetic mixing between the photon and dark photon, as they couple to both a photon and a dark photon.
If the dark charges are of $O(1)$, we expect the induced kinetic mixing to be about $O(10^{-1})$ for the UV scale being the typical grand unification scale of  $10^{16}$ GeV and the exotic fermion masses at the TeV scale \cite{Kaneta:2017wfh}.
It does not mean we cannot take zero kinetic mixing.
The kinetic mixing at the UV scale is not determined in general, and we can fine-tune its value such that the net value after the loop correction is negligibly small at the EW scale.  
Alternatively, we can take the approaches in Refs.~\cite{Holdom:1985ag,Davoudiasl:2012ag}, which assume another set of exotic fermions with equal and opposite dark charges to cancel the contributions of the two sets of exotic fermions.

\subsection{Constraints}
There are several experimental constraints to consider from Higgs fits, EW precision data, Higgs searches at LEP, and heavy scalar searches at the LHC. First, all of our signals will modify the total Higgs width. Namely, they contribute to the unknown branching fractions of the SM Higgs. The current $95\%$ C.L. upper limit on the unknown branching fraction from the LHC is $0.12$ \cite{ATLAS:2022vkf}.  Secondly, all the above constraints place limits on how large the scalar mixing angle can be.  For scalar masses less than about 100 GeV, Higgs searches at LEP limit the mixing angle to $|\sin \theta | \alt 0.1$ \cite{Robens:2015gla}. For dark Higgs masses in the 200 GeV to 1 TeV range, the mixing angle is limited to $|\sin \theta | \alt 0.1-0.2$ from EW precision data, Higgs fits, and scalar searches at the LHC~\cite{Adhikari:2020vqo, Robens:2022oue, Adhikari:2022yaa}.

As discussed above, in the dark Higgs scenario, the coupling $\lambda_{h\gamma_D\gamma_D}$ is given by Eq.~(\ref{eq:HiggsDarkPhoton}). Limits on the scalar mixing angle will then place an upper bound on the branching ratio of the Higgs to two dark photons:
\begin{align}
\text{BR}(h\rightarrow \gamma_D\gamma_D)\leq \text{BR}^{\text{max}}( h \rightarrow \gamma_D \gamma_D ) & \approx
\frac{\Gamma^{\text{max}}( h \rightarrow \gamma_D \gamma_D )}{\Gamma_\text{SM}(h) + \Gamma^{\text{max}}( h \rightarrow \gamma_D \gamma_D )},
\label{eqn:BRmax}
\end{align}
where $\Gamma_\text{SM}(h) = 4.10 $ MeV \cite{LHCHiggsCrossSectionWorkingGroup:2013rie} is the SM value of the total width, $m_h = 125.18$ GeV \cite{ParticleDataGroup:2018ovx}  is the Higgs mass.
The maximum width is
\begin{align}
\Gamma^{\text{max}}( h \rightarrow \gamma_D \gamma_D ) & =
\frac{|\lambda_{h\gamma_D\gamma_D}^{\text{max}}|^2}{128 \pi m_h} \left(12 - 4 \frac{m_h^2}{m_{\gamma_D}^2} + \frac{m_h^4}{m_{\gamma_D}^4}\right)
\left(1 - 4 \frac{m_{\gamma_D}^4}{m_h^4}\right)^{\frac{1}{2}},
\end{align}
and the maximum coupling is
\begin{eqnarray}
|\lambda_{h\gamma_D\gamma_D}^{\text{max}}| \sim |\sin \theta^{\text{max}}| \frac{m_{\gamma_D}^2}{v_D} \approx 0.1 \frac{m_{\gamma_D}^2}{v_D}.
\end{eqnarray}
We have assumed that the BR($h \rightarrow a a $) is negligible.

\begin{figure}[tb]
\begin{center}
\subfigure{\includegraphics[width=0.49\textwidth, clip]{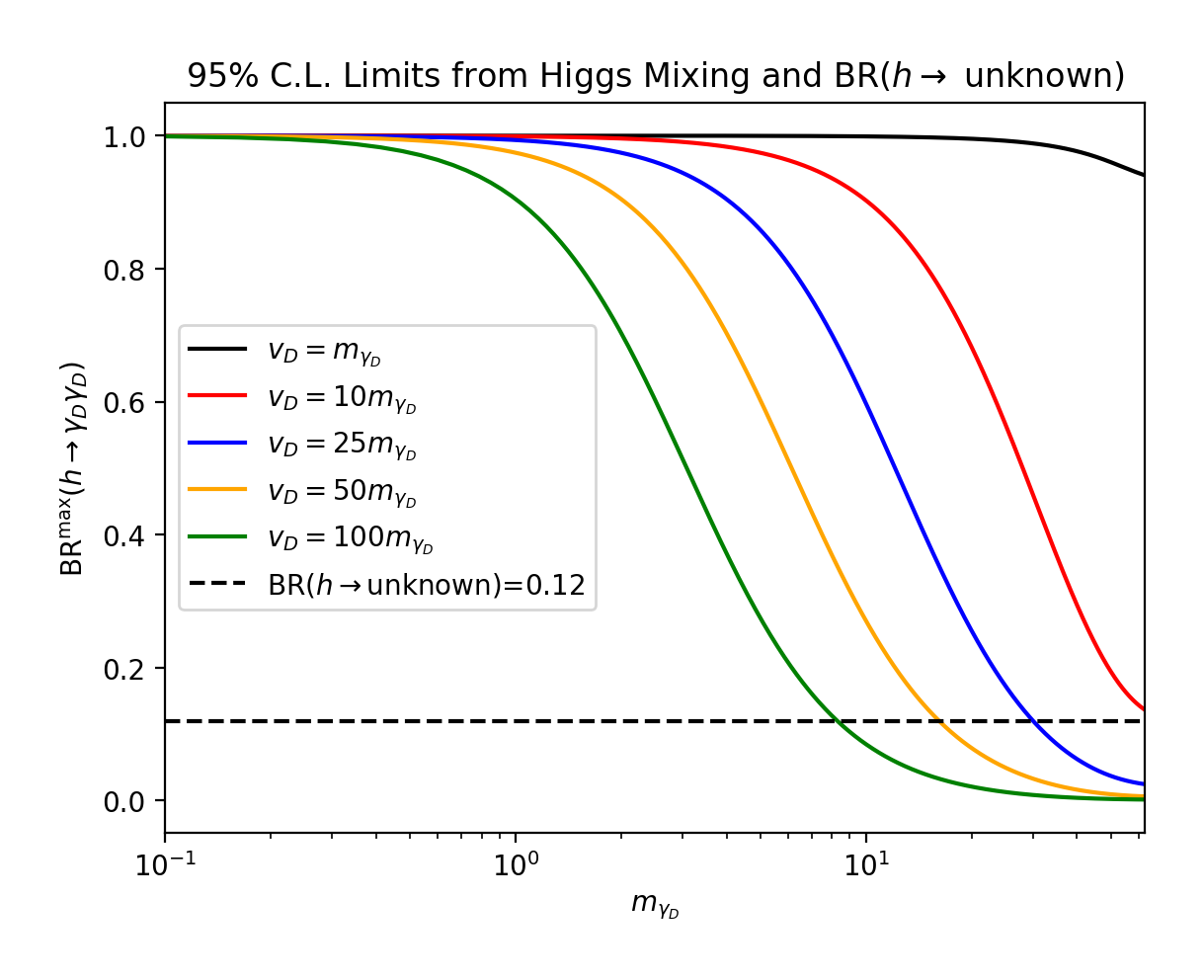}}
\end{center}
\caption{
\label{fig:BRmax} Maximum branching ratio of the Higgs to two dark photons from (dashed) Higgs fits to the branching fraction of Higgs to the unknown and from (solid) Higgs mixing assuming $\sin\theta^{\rm max} = 0.1$ with (black) $v_D= m_{\gamma_D}$, (red) $v_D=10\,m_{\gamma_D}$, (blue) $v_D=25\,m_{\gamma_D}$, (yellow) $v_D=50\,m_{\gamma_D}$, and (green) $v_D=100\,m_{\gamma_D}$.}
\end{figure}

The maximum branching ratio of Higgs to two dark photons for different ratios of $m_{\gamma_D}/v_D$ and the maximum branching ratio from Higgs to the unknown limits are shown in Fig.~\ref{fig:BRmax}. We see that the scalar mixing angle bound is not really relevant until $v_D$ is greater than about 10 times $m_{\gamma_D}$. For $v_D = 100\,m_{\gamma_D}$, these constraints surpass the limits on Higgs branching ratio to unknown for $m_{\gamma_D}\gtrsim 10$~GeV. The sensitivity falls off at low masses, since the decays of $h\rightarrow \gamma_D\gamma_D$ are dominated by the longitudinal modes.  In this limit, 
\begin{eqnarray}
\Gamma^{\rm max}(h\rightarrow\gamma_D\gamma_D)\xrightarrow[m_{\gamma_D}\rightarrow 0]{} \frac{|\sin\theta^{\rm max}|^2}{128\,\pi}\frac{m_h^3}{v_D^2}.
\end{eqnarray}
Hence, as $m_{\gamma_D}$, and subsequently $v_D$, decreases, the maximum branching ratio approaches one for fixed $\sin\theta^{\rm max}$. Then the limit on the branching ratio to undetected final states is the most relevant.

There are also constraints on the ALPs couplings to both the photons and dark photons coming from astrophysical measurements, beam-dump experiments, dark photon conversions, mono-photon searches, and heavy colored fermions searches \cite{Dobrich:2019dxc, Lucente:2022jxm,Carenza:2020zil,deNiverville:2018hrc, Deniverville:2020rbv, Lanfranchi:2020crw,CMS:2018erd,ATLAS:2020hii,Aloni:2019ruo,Belle-II:2020jti,Knapen:2016moh,Lee:2018lcj}.  For $m_a\sim 6-62.5$~GeV, ultraperipheral lead-lead collisions at the LHC place a limit of $G_{a\gamma\gamma}\lesssim 5\times10^{-5}-1\times10^{-4}~{\rm GeV}^{-1}$~\cite{ATLAS:2020hii,CMS:2018erd}. One of the most relevant constraints for $m_a\sim 0.04-6$~GeV are the LEPII bounds of $G_{a\gamma\gamma}\lesssim 10^{-3}~{\rm GeV}^{-1}$~\cite{Knapen:2016moh}. In the mass range $m_a\sim 0.1-1$~GeV, a photon beam experiment~\cite{Aloni:2019ruo} and Belle II searches~\cite{Belle-II:2020jti} are relevant, making the most stringent constraints $G_{a\gamma\gamma}\lesssim 7\times10^{-4}-1\times10^{-3}~{\rm GeV}^{-1}$.  Below $m_a\sim0.04$ GeV, the beam dump experiments limit the coupling  $G_{a\gamma\gamma} \alt 10^{-7}~\text{GeV}^{-1}$ \cite{Dobrich:2019dxc}, and constraints from SN1987 force $G_{a\gamma\gamma}\lesssim 10^{-10}-10^{-9}~{\rm GeV}^{-1}$ \cite{Lee:2018lcj}. For the dark axion portal coupling, $G_{a\gamma\gamma_D} \lsim 0.002~\text{GeV}^{-1}$ was obtained for $m_{\gamma_D} \agt 0.1$ GeV when $G_{a\gamma\gamma}$ is turned off and $m_a \ll m_{\gamma_D}$. We will use this value for the representative bound in our numerical study. \cite{deNiverville:2018hrc, Deniverville:2020rbv}. Depending on the available decay channels, such small couplings can lead to long-lived ALPs and dark photons. In our scenario, the only available decay of the ALP is to photons. The only available decay of the dark photon is to a photon and an ALP. Thus, we take their branching ratios to be one. We will consider the implications of the coupling limits on decay length below.

\subsection{Production and Decay Rates}
This work will assume all decays are prompt and on-shell. To ensure on-shell decays, we place an upper bound of $m_h/2 \approx 62.5$ GeV on the dark photon mass. We also require the ALP mass to be less than the dark photon mass.
We take the Higgs on shell and use the narrow width approximation (NWA) on the propagators. This separates the production and decay at the total cross section level, as 
\begin{align}
\sigma( p p \rightarrow h \rightarrow \gamma_D \gamma_D \rightarrow 6 \gamma ) & \approx  
\sigma(p p \rightarrow h) 
\text{BR} \left( h \rightarrow \gamma_D \gamma_D \right)
\text{BR}^2 \left(\gamma_D \rightarrow a \gamma \right)
\text{BR}^2 \left( a \rightarrow \gamma \gamma \right) .
\end{align}

One should be careful that the ALP and dark photon decays are in fact prompt. This effectively places a lower bound on the dark photon and ALP masses.  The primary decay channel for ALPs will be to photon pairs. The on-shell, spin-averaged decay width for the ALP is given by
\begin{align}
\Gamma(a \rightarrow \gamma \gamma) & = \frac{|G_{a\gamma \gamma}|^2 m_a^3}{64 \pi} .
\end{align}
If we assume $ G_{a\gamma\gamma}  \approx 10^{-3}~\text{GeV}^{-1} $,  for an ALP of mass $40$~MeV the rest frame decay width will be about $3\times 10^{-13}$~GeV and the rest frame decay length is $0.62$ mm.  
The primary decay channel for dark photons in the zero kinetic mixing limit will be to an ALP and a photon. The on-shell, spin-averaged decay width is given by
\begin{align}
\Gamma(\gamma_D \rightarrow \gamma a)  &= 
\frac{ |G_{a\gamma \gamma_D}|^2m_{\gamma_D}^3}{96 \pi} \left(1 - \frac{m_a^2}{m_{\gamma_D}^2}\right)^3.
\end{align}
Assuming $G_{a\gamma\gamma_D}\approx 2\times10^{-3}~{\rm GeV}^{-1}$, $m_{\gamma_D}=0.1~{\rm GeV}$, and $m_a=40~{\rm MeV}$, the dark photon decay width is about $8\times 10^{-12}~{\rm GeV}$ and the rest frame decay length is $25~\mu{\rm m}$.

\begin{figure}[tb]
\subfigure[]{\includegraphics[width=0.45\textwidth,clip]{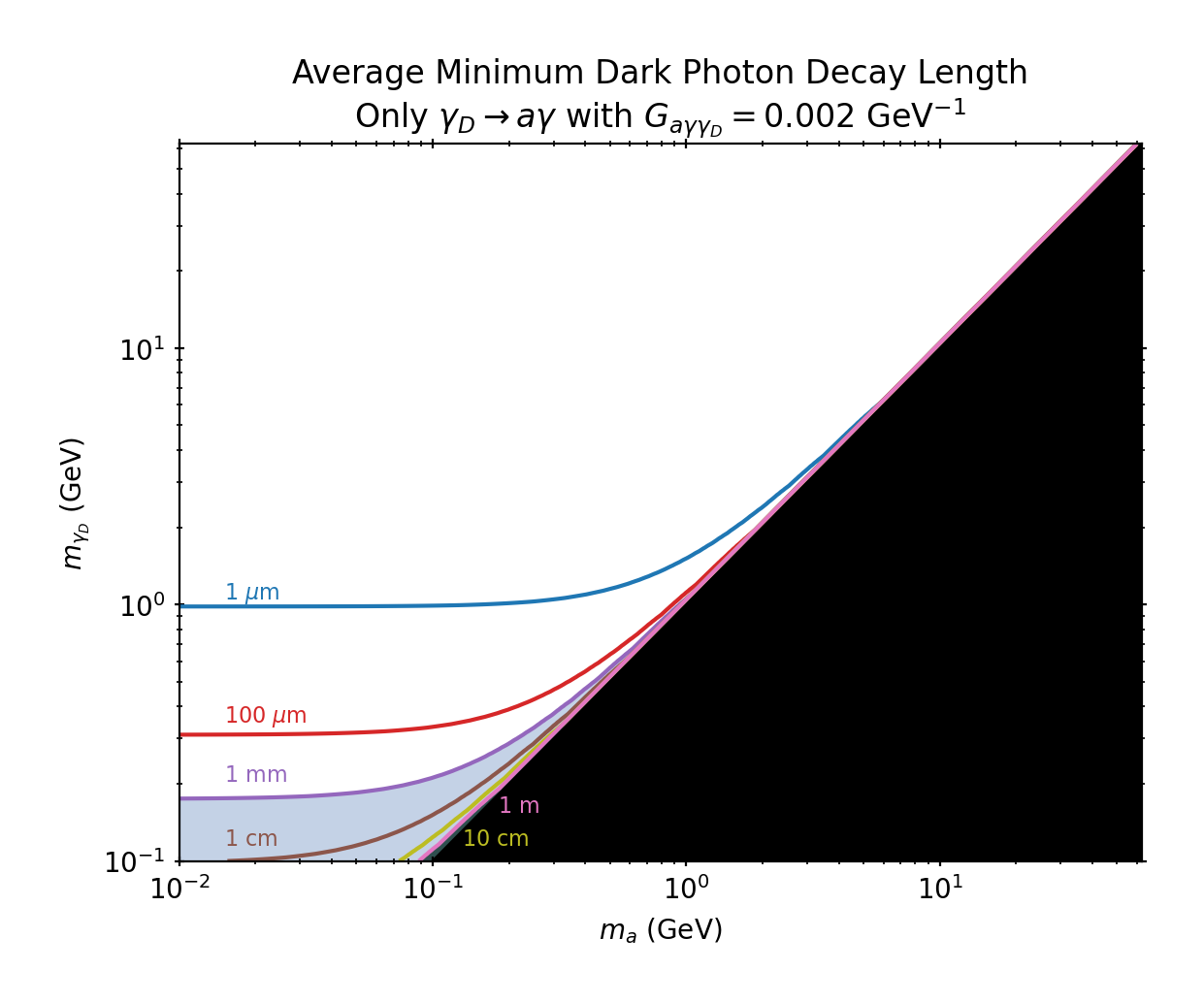}}
\subfigure[]{\includegraphics[width=0.45\textwidth,clip]{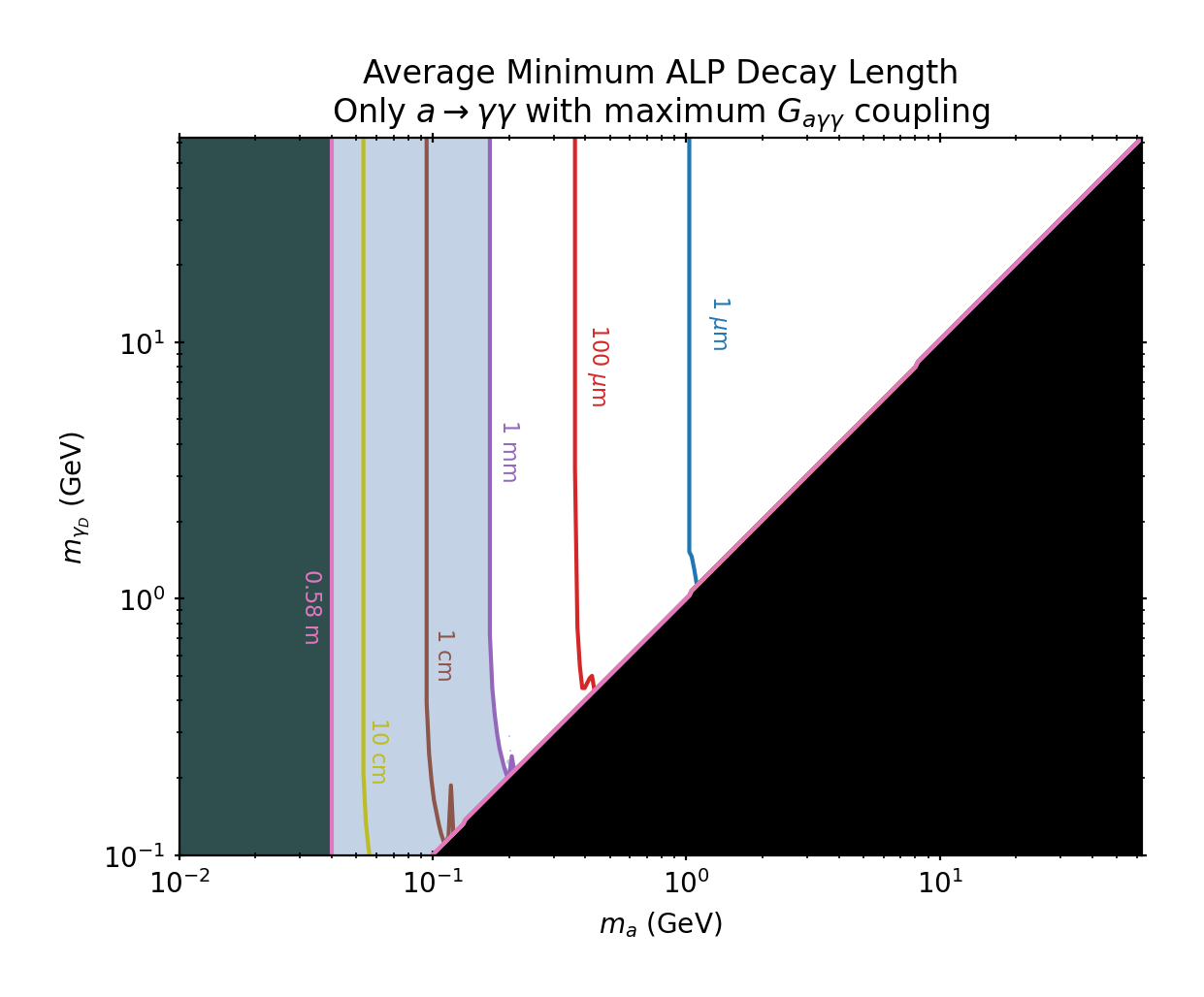}}
\caption{\label{fig:DecayLengths} Average decay lengths for the (a) dark photon and (b) ALP in the Higgs rest frame.  Blue is for decay lengths of $1~\mu$m, red for $100~\mu$m, purple for $1$~mm, brown for $1$~cm, yellow for $10$~cm, and magenta for $1$~m is the dark photon case and $0.58$~m for the ALP case.}
\end{figure}

While prompt on detector lengths, the above decay lengths are calculated in the rest frame of the parent particle.  However, these decay lengths can be considerably lengthened by boosts to the Higgs rest frame.  In Fig.~\ref{fig:DecayLengths}, we show the average decay lengths of the (a) dark photon and (b) ALP in the Higgs rest frame.  The average is calculated including the matrix element weight (see App.~\ref{sec:BR}).  Additionally, $G_{a\gamma\gamma}$ and $G_{a\gamma\gamma_D}$ are set to their maximum allowed values.  As can be seen for dark photons with masses $m_{\gamma_D}\gtrsim 0.3$~GeV, the decays are indeed prompt.  For dark photon masses between 0.1 and 0.3~GeV, the average decay length is displaced or inside the tracker~\cite{Liu:2015bma,Alimena:2019zri,Alimena:2021mdu,CMS:2014wda,ATLAS:2015oan,CMS:2016vuw,CMS:2017poa,ATLAS:2017tny,ATLAS-CONF-2017-026}.  When $m_{\gamma_D}$ and $m_a$ are comparable, the phase space for the dark photon decay is squeezed increasing the decay lengths in this region.  Additionally, the ALPs are prompt for masses $m_a\gtrsim 0.4$~GeV and decay is displaced and/or within the tracker for masses between $0.04$ and $0.4$~GeV. The spikes near the threshold $m_a\approx m_{\gamma_D}$ for the lines of constant ALP decay length are a result of the coupling strength limits from the photon beam and Belle II.  For ALP masses $m_a\lesssim 0.04$~GeV, the constraints on $G_{a\gamma\gamma}$ become considerably stronger. Hence, the average ALP decay length is outside the LHC detectors.  

While some average decay lengths are not necessarily prompt, they are within the tracker for $m_a\gtrsim 0.04$~GeV and $m_{\gamma_D}\gtrsim 0.1$~GeV. As we will show, the mass regions with longer decay lengths have a large overlap with signals that appear as two or four $\gamma$-jets. To observe the $\gamma$-jets to be different from single photons, it would be necessary to determine that there is a displaced vertex.  Since photons do not leave tracks in the tracker, it is difficult to observe the displaced vertex in the $\gamma$-jets.  However, recently there has been a proposal to use photon conversions in the ATLAS tracker to detect displaced vertices in diphoton ALP decays~\cite{Alonso-Alvarez:2023wni}.  The largest irreducible backgrounds to the displaced ALP decay are displaced $K_L$ decays.  The authors of Ref.~\cite{Alonso-Alvarez:2023wni} consider a background free region where the ALPs have transverse momentum above $\sim120$~GeV.  Since the parent particles of the displaced vertices under consideration in this paper originate from Higgs decays, their transverse momenta are of the order of 10~GeV.  Hence, the $K_L$ background is important.   In this region, the relevant $K_L\rightarrow \gamma\gamma$ cross section, including the probability of displaced vertices, is of the order of $10^{-2}$~pb~\cite{Alonso-Alvarez:2023wni}.  Searches for four-photon Higgs decays place an upper bound on the cross section of $\mathcal{O}(10^{-3})$~pb~\cite{CMS:2022xxa}.  Current measurements of the Higgs to diphoton limit additional contributions to the diphoton rate to be at most $\mathcal{O}(10^{-2})$~pb~\cite{CMS:2021kom, ATLAS:2022tnm}.  It would be interesting to determine if the displaced vertices of the signal could be distinguished from the $K_L$ backgrounds.  However, since the signal cross section is at best comparable to the $K_L$ backgrounds, for the purposes of this study we will assume the displaced vertices of our signal cannot be distinguished and treat $\gamma$-jets as single photons.

Now, in order to estimate the cross sections and therefore rates, we need the Higgs production cross section, 
$\text{BR}\left(h \rightarrow \gamma_D \gamma_D \rightarrow a \gamma a \gamma \right)$, and $\text{BR}\left(a \rightarrow \gamma \gamma\right)$.  
Assuming on shell intermediate particles and using phase space recursion, the branching ratios can be written as 
\begin{align}
\text{BR}\big( &h \rightarrow \gamma_D(k_1) \gamma_D(k_2) 
\rightarrow \gamma(p_1) a(q_1) \gamma(p_2) \gamma(q_2) 
\rightarrow \gamma(p_1) \gamma(p_3) \gamma(p_4) \gamma(p_2) \gamma(p_5) \gamma(p_6) \big)\nonumber
\\ &=
\text{BR}(h\rightarrow \gamma_D\gamma_D)
~\text{BR}(\gamma_D\rightarrow a \gamma)^2 
~\text{BR}(a\rightarrow \gamma\gamma)^2 
\frac{9
\int d\Omega_h d\Omega_{\gamma_D(k_1) } d\Omega_{\gamma_D(k_2)}f(p_1, k_1, p_2, k_2) }{ (2 \pi)^3 \left(12 - 4 \frac{m_h^2}{m_{\gamma_D}^2} + \frac{m_h^4}{m_{\gamma_D}^4}\right)  (m_{\gamma_D}^2 - m_a^2)^4 } ,
\label{eq:ampSq}
\end{align}
where the momenta correspond to those in Fig.~\ref{fig:sixphoton}.  The solid angles $\Omega$ are the phase space variables for the respective two-body decays of each parent particle with the subscript indicating the parent particle.  The function $f(p_1, k_1, p_2, k_2)$ is given by
\begin{align}
f(p_1, k_1, p_2, k_2) & = (p_2 \cdot k_2 ) (k_1 \cdot k_2 ) (p_1 \cdot p_2 ) (p_1 \cdot k_1 ) 
+ (p_2 \cdot k_1 ) (p_2 \cdot k_2 ) (p_1 \cdot k_2 ) (p_1 \cdot k_1 ) \nonumber
\\ & \ \ \ \ \ \ \ \ \ \ \ \ \ 
- (p_2 \cdot k_1 ) (p_1 \cdot p_2 ) (p_1 \cdot k_1 ) k_2^2
- (p_2 \cdot k_2 ) (p_1 \cdot k_2 ) (p_1 \cdot p_2 ) k_1^2 \nonumber
\\ & \ \ \ \ \ \ \ \ \ \ \ \ \  
+ \frac{1}{2}(p_1 \cdot p_2 )^2 k_1^2 k_2^2 .
\label{eq:f}
\end{align}
This function encapsulates the spin correlations of the decay chain.  Since $a \rightarrow \gamma \gamma$ is isotropic, $f$ is independent of the ALP decay products momenta and the respective phase space integrations are trivial.

Additionally, we can ignore the vector-boson fusion (VBF) of the Higgs from dark photons, since they do not couple to the SM particles in the negligible kinetic mixing limit. The total Higgs production cross section will be scaled by the Higgs mixing angle $\theta$ as $\sigma(p p \rightarrow h) = \cos^2 \theta \sigma_\text{SM}(p p \rightarrow h) $. In the small $\theta$ limit, one can approximate the Higgs production rate as the SM rate. In the following, we use the inclusive cross section at the 13 TeV LHC~\cite{LHCHiggsCrossSectionWorkingGroup:2016ypw}
\begin{eqnarray}
\sigma_\text{SM}(gg\rightarrow h)=52~{\rm pb}.\label{eq:xsect}
\end{eqnarray}
To accurately simulate final state particle rapidities, we use the gluon-gluon parton luminosity
\begin{eqnarray}
\mathcal{L}(\tau)=\int_{\ln\sqrt{\tau}}^{-\ln\sqrt{\tau}}dy_h g(\sqrt{\tau}e^{y_h},\mu_F)g(\sqrt{\tau}e^{-y_h},\mu_F),
\label{eq:partlum}
\end{eqnarray}
where $\tau=m_h^2/S$, $\sqrt{S}=13$~TeV is the LHC lab frame energy, $\mu_F=m_h/2$ is the factorization scale, and $y_h$ is the Higgs rapidity.  We use \texttt{LHAPDF}~\cite{Buckley:2014ana} with the \texttt{CTEQ18NLO} \cite{Hou:2019efy} parton distribution functions.

\section{Signal Categories}
\label{sec:sig}

To classify our signal, we need the relative signal rates of each possible multi-photon category. For the rest of the paper we only consider $m_a\gtrsim 0.04$~GeV and $m_{\gamma_D}\gtrsim0.1$~GeV such that the dark photon and ALP decays occur before the ECAL.  We classify the multi-photon categories according to the number of ``isolated photons'' (isolated single photons and photon-jets) and $\xi$-jets.  To do this, we generate events and event weights by sampling phase space according to the function $f$ in Eq.~(\ref{eq:f}) and isotropic decays in the ALP rest frame, and Higgs rapidity according to the parton luminosity in Eq.~(\ref{eq:partlum}).
 The signal categories are classified using the angular separation between photons:
\begin{eqnarray}
\Delta R_{ij}=\sqrt{(\Delta \eta_{ij})^2+(\Delta \phi_{ij})^2},
\end{eqnarray}
where $\Delta\eta_{ij}$ and $\Delta\phi_{ij}$ correspond to the rapidity and azimuthal angle differences between two particles $i$ and $j$.  As described below, we first construct photon-jets.   The photon-jets and photons are the so-called ``observed photons.''   At this stage, we can separate our signals into categories based on the number of observed photons.  However, these observed photons may not be isolated.  After constructing photon-jets, we implement isolation criteria and construct $\xi$-jets. Then we can create categories based on the number of isolated photons and the number of $\xi$-jets in an event.

\begin{figure}[tb]
\begin{center}
\subfigure{\includegraphics[width=0.49\textwidth, clip]{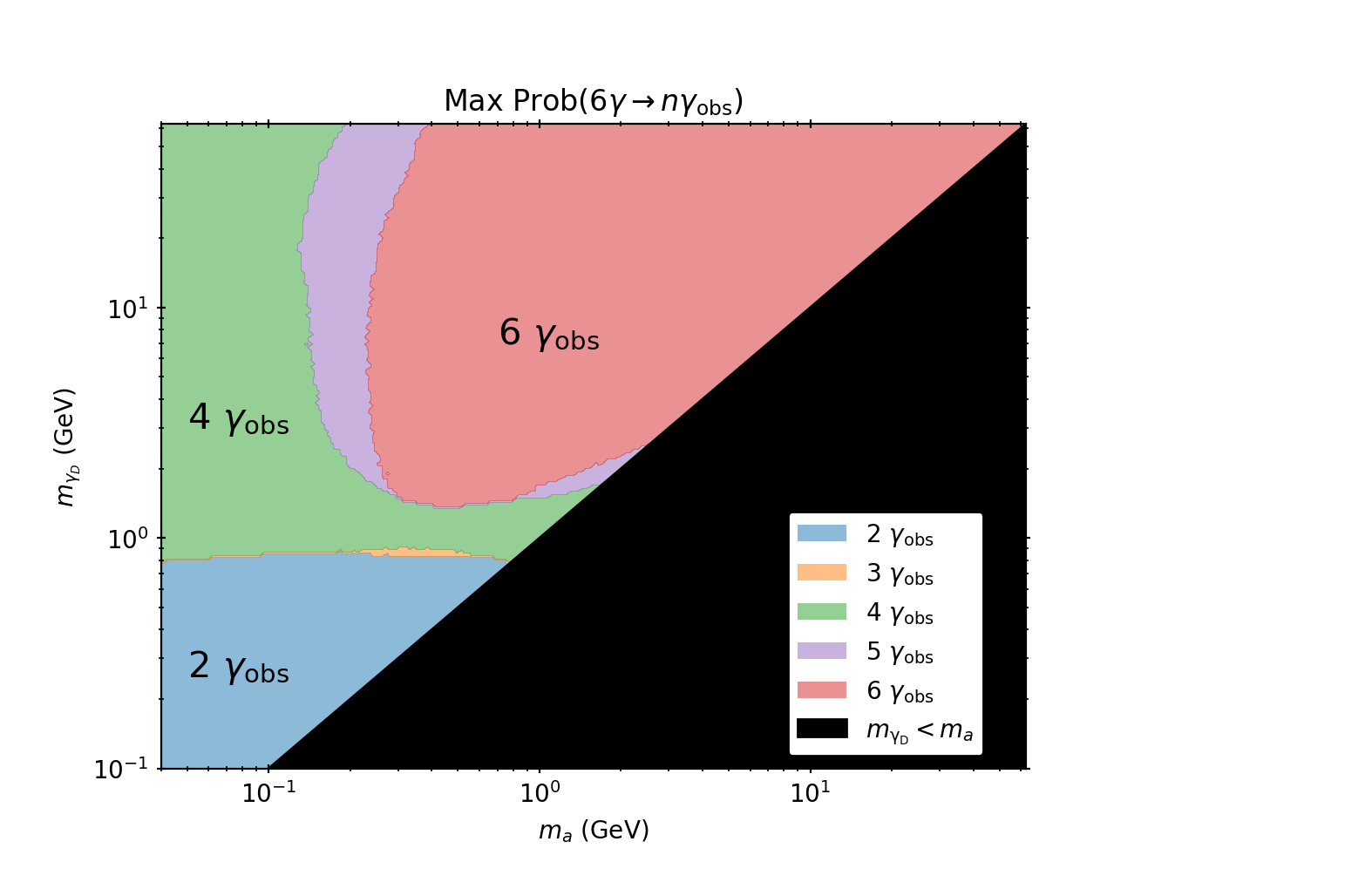}}
\end{center}
\caption{
\label{fig:DomBRMerge} 
Regions with the largest probability of ending in $n$ observed photon final states after merging photons into $\gamma_{\rm obs}$: 
(blue) two, (orange) three, (green) four, (purple) five, and (red) six observed photons.  No isolation requirements are imposed.  The black corner indicates the kinematically forbidden area ($m_{\gamma_D} < m_a$).}
\end{figure}

\subsection{$\gamma-$Jets}

We define $\gamma$-jets as having an angular separation  $\Delta R_{ij}\leq 0.04$ \cite{Sheff:2020jyw}, such that the photons would appear in a single ECAL cell. 
Photon-jets are treated as indistinguishable from single photons, and we label both as observable photons ($\gamma_{obs}$) to emphasize they may not all be single photons. No isolation requirements are imposed on the observable photons.  The procedure to construct photon-jets from truth-level photons is as follows:
\begin{enumerate}
\item{Generate truth-level photon momentum.}
\item{Merge photons into photon-jets:}
    \begin{enumerate}
        \item Calculate angular separation $\Delta R_{ij}$ between all photon objects, where $i$ and $j$ label the photon objects.
        \item If the smallest $\Delta R_{ij} \leq 0.04$, then remove $i$th and $j$th photon momentum and add a new photon object with momentum $p = p_i + p_j$. 
        
        \item Repeat the above steps 2(a) and 2(b) until no photons are merged.
        \item The remaining objects are the $\gamma_{obs}$ discussed above.
    \end{enumerate}
\end{enumerate}

Figure~\ref{fig:DomBRMerge} shows the regions in the $m_a$-$m_{\gamma_D}$ plane with the largest probability into (blue) two, (orange) three, (green) four, (purple) five, and (red) six observed photons. 
First, on-shell decay conditions require $m_{\gamma_D} > m_a$. The kinematically excluded region, $m_{\gamma_D}<m_a$, is shown in black. Near the black edge, the ALP is produced nearly at rest and the photons coming from the dark photon are exceptionally soft. 

In the top right region, the masses $m_a \sim m_{\gamma_D} \sim m_h/2$ and both the dark photon and ALP are produced nearly at rest. Here the resulting photons do not receive much boost and are well separated. Thus, the dominant branching fraction is to six-photon events. In the top left region, $m_a \ll m_{\gamma_D}$ and $m_{\gamma_D} \sim m_h/2$. The ALP receives a large boost, and the photons coming from the ALP are merged into photon-jets. However, the dark photon is significantly less boosted and the ALP and photon originating from the dark photon are well-separated. Hence, the dominant branching ratio is to four-photon events (2$\gamma$'s+2$\gamma$-jets).  When $ m_{\gamma_D}  \ll m_h/2$, the dominant branching ratio is to two-photon events, since all decay products from the dark photons are well-collimated.  There are also some small transition regions where five-photon and three-photon decays are favored. 

One can estimate when the different signal categories become dominant by calculating the peak of the $\Delta R$ in the various limits. First, we calculate the turnover between the six and four-photon signals.  
When  $m_a \ll m_{\gamma_D} \approx m_h/2$ the dark photon is nearly at rest and the ALP can be treated as massless.  Hence, a dark photon energy, $m_h/2$, is transferred equally to its decay products.  The ALP's energy is then given by $E_a\approx m_h/4$. The angular separation of the two photons from the ALP decay can be calculated as
\begin{align}
\Delta R_{\gamma \gamma} \approx 2 \frac{m_{\gamma_a}}{ E_{a} }\approx 8 \frac{m_a}{m_h}.
\end{align}
Therefore, the angular separation between the ALP decay products $\Delta R_{\gamma \gamma} \leq 0.04$  when
\begin{align}
  m_a \alt \frac{m_h}{200} \approx 0.625~\text{GeV}.
\end{align} 

When $m_a\approx m_{\gamma_D}$, then the ALP is produced nearly at rest in the dark photon rest frame ($E_a\approx m_a$) with a very soft photon ($E_\gamma\approx 0$).  In the Higgs rest frame, the ALP energy is comparable to that of the dark photon: $E_a\approx m_h/2$.  Hence, the angular separation between the ALP decay products is
\begin{eqnarray}
\Delta R_{\gamma\gamma}\approx 4\frac{m_a}{m_h}.
\end{eqnarray}
The condition $\Delta R_{\gamma\gamma}\leq 0.04$ is then satisfied when
\begin{eqnarray}
m_{\gamma_D}\approx m_a\lesssim \frac{m_h}{100}\approx 1.25~{\rm GeV}.
\end{eqnarray}
While the ALP decay products are collimated into a $\gamma$-jet, the photon and ALP from the $\gamma_D$ are not necessarily well-collimated.  
Hence, we expect the transition between the six and four observed photons to be when $m_{a}\sim 0.625$~GeV or $m_{\gamma_D}\sim 1.25$~GeV, as seen in Fig.~\ref{fig:DomBRMerge}.

For $m_{\gamma_D}\ll m_h$, events have two observed photons.  That is, the angular separation between all three photons from a dark photon decay must be collimated. To make this concrete, label the photons as $\gamma_D\rightarrow \gamma_1 a\rightarrow \gamma_1\gamma_2\gamma_3$, where $\gamma_2,\gamma_3$ are from ALP decays.  Then we need
\begin{eqnarray}
{\rm max}(\Delta R_{\gamma_1\gamma_2},\Delta R_{\gamma_1\gamma_3}, \Delta R_{\gamma_2\gamma_3})\leq 0.04.\label{eq:DPdecays}
\end{eqnarray}
The configuration where the photons have the largest separation is when the dark photon and ALP decay planes align.  In this limit, we first consider the angular separation between $\gamma_1$ and the ALP decay products:
\begin{eqnarray}
{\rm max}(\Delta R_{\gamma_1\gamma_2},\Delta R_{\gamma_1\gamma_3})\approx \Delta R_{\gamma_1 a}+\frac{\Delta R_{\gamma_2\gamma_3}}{2}\leq 0.04.
\end{eqnarray}
A necessary condition of Eq.~(\ref{eq:DPdecays}) is $\Delta R_{\gamma_2\gamma_3}\leq 0.04$, which implies
\begin{eqnarray}
\Delta R_{\gamma_1a}\approx \frac{2\,m_{\gamma_D}}{E_{\gamma_D}}\lesssim 0.02.
\end{eqnarray}
Using $E_{\gamma_D}=m_h/2$, this leads to the condition
\begin{eqnarray}
m_{\gamma_D}\lesssim \frac{m_h}{200}\approx 0.625~{\rm GeV}.
\end{eqnarray}
We note that since $m_a\leq m_{\gamma_D}$, this condition is consistent with $\Delta R_{\gamma_2\gamma_3}\lesssim 0.04$.  Hence, for $m_{\gamma_D}\lesssim 0.625$~GeV, we expect the signal to have two observed photons.  This is precisely what is seen in Fig.~\ref{fig:DomBRMerge}.

\begin{figure}[tb]
\begin{center}
    \subfigure{\includegraphics[width=0.49\textwidth,clip]{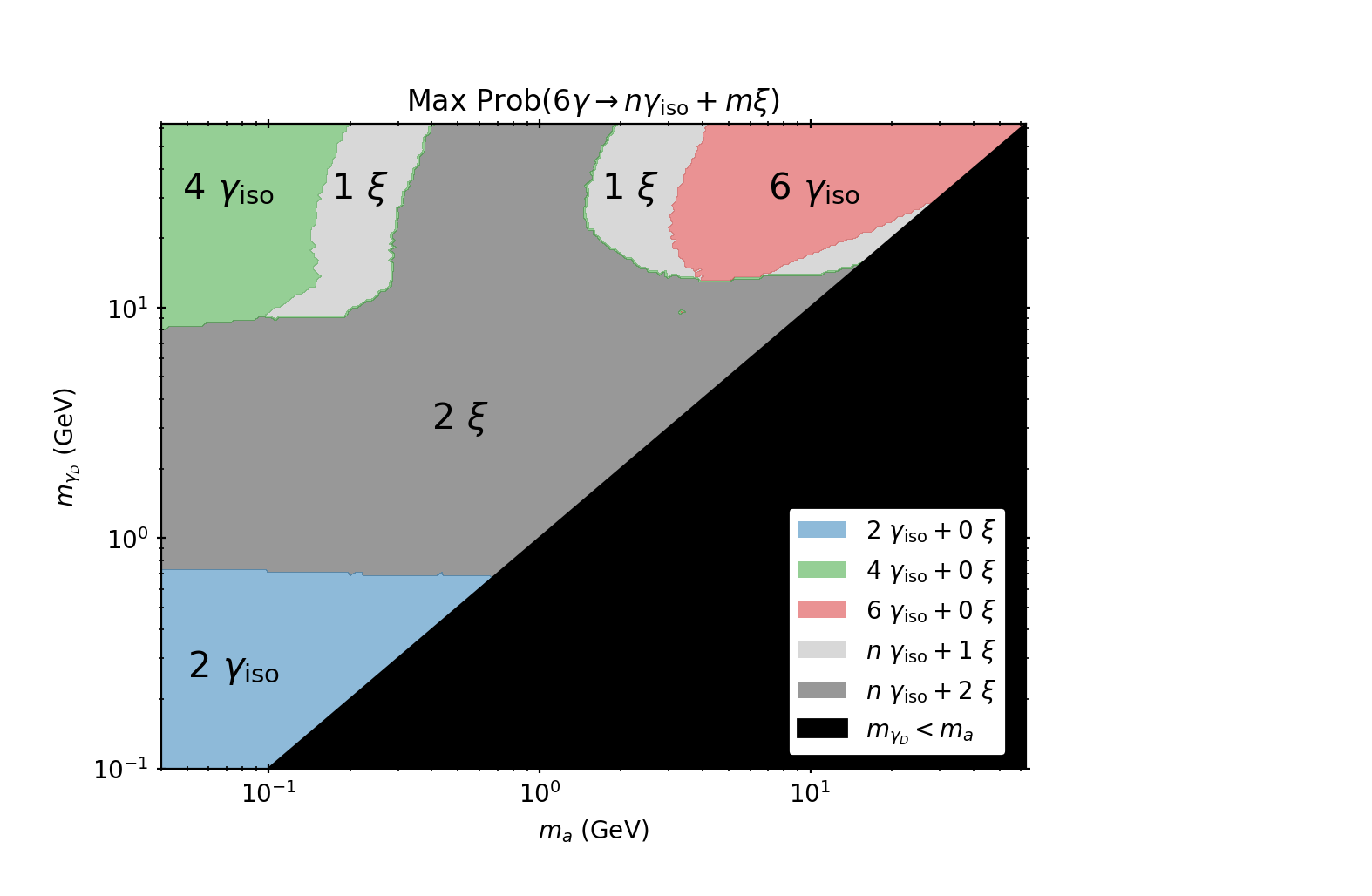}}
\end{center}
\caption{
\label{fig:DomBRIso}
Regions with the largest probabilities into $n$ isolated photons and $m$ $\xi$-jet final states.  The light gray (dark gray) region indicates the dominant final states have one (two) $\xi$-jets. Color coding for other regions: (blue) $2\gamma_{\rm iso}+0\xi$-jet, (green) $4\gamma_{\rm iso}+0\xi$-jet, and (red) $6\gamma_{\rm iso}+0\xi$-jet.  }
\end{figure}

\subsection{$\xi-$Jets}

Experiments typically require photons to be well-isolated in order to help eliminate QCD backgrounds. As such, we require signal photons to be isolated in a cone of $\Delta R < 0.4$. This means any photon pairs with $\Delta R$ between the photon-jet minimum and the isolation requirement will not be identified as photons in the detector. 
The convention of defining $\xi$-jets to be multi-photon signals with angular separation $0.04 <\Delta R < 0.4$ is adopted~\cite{Sheff:2020jyw}. 

After steps 1 and 2 of the merging procedure described above, we isolate photons and construct $\xi$-jets via step 3 below:
\begin{enumerate}
\item[3.]{Isolate photons and construct $\xi$-jets from non-isolated objects.}
    \begin{enumerate}
        \item Calculate angular separation $\Delta R$ between all $\gamma_{\rm obs}$ and $\xi$-jets. 
        \item If the smallest $\Delta R_{ij} < 0.4$, remove the $i$th and $j$th observed photons or $\xi$-jets and combine them into a new $\xi$-jet with momentum $ p = p_i + p_j.$ 
        \item Repeat steps 3(a) and 3(b) until no more $\xi$-jets are formed.
        \item Any observed photons not combined into $\xi$-jets are identified as the isolated photons $\gamma_{iso}$.
    \end{enumerate}
\end{enumerate}
In the following, we define the probability of $n$ isolated photons and $m$ $\xi$-jets as 
 \begin{align}
\text{Prob}(6\gamma \rightarrow n\gamma_{iso} + m \xi) & = 
 \frac{\text{BR}(h \rightarrow \gamma_D \gamma_D \rightarrow a \gamma a \gamma \rightarrow 6 \gamma \rightarrow n\gamma_{iso} + m \xi)}
 {\text{BR} \big( h \rightarrow \gamma_D \gamma_D\big)
\text{BR}^2 \big(\gamma_D \rightarrow a \gamma\big)
\text{BR}^2 \big( a \rightarrow \gamma \gamma \big)}.
\label{eq:kinBR}
 \end{align}

The effects of isolating photons and constructing $\xi$-jets can be seen in Fig.~\ref{fig:DomBRIso}. 
The figure shows regions with the largest probability into different signal categories. Again, for on-shell decays the black region is excluded since  $m_a > m_{\gamma_D}$ in that region.  One can see a structure similar to that of Fig.~\ref{fig:DomBRMerge}. 
The major difference is that much of the parameter space results in either (light gray) one $\xi$-jet or (dark gray) two $\xi$-jets. In these regions, the dark photons and/or ALPs have intermediate boosts. Hence, the observed photons are sufficiently collimated to fail isolation requirements and form $\xi$-jets. This region largely consists of $0\gamma_{\rm iso}+2\xi$-jets, $3\gamma_{\rm iso}+1\xi$-jet,  $4\gamma_{\rm iso}+1\xi$-jet, and $2\gamma_{\rm iso}+2\xi$-jets final states.  The regions dominated by $\xi$-jets could be shrunk by decreasing the isolation requirement on $\Delta R$. However, independent of relaxing isolation requirements, we will show that it is possible some of these events have been saved on tape.

\section{Results}
\label{sec:result}

Our signals of two and four isolated photons with zero $\xi$-jet have been searched for at the LHC~\cite{ATLAS:2022tnm,CMS:2021kom,CMS:2022xxa}.  In this section, we recast those results into constraints on our signal of $h\rightarrow 6\gamma$.  In the process, we will discuss the detector efficiencies of our signal process.  This analysis also motivates new searches for six-isolated photon signals and signals with $\xi$-jets.  We will also show that many events with $\xi$-jets may very well have passed the two or three-photon triggers.  We note that even if an event with six-isolated photons or non-zero $\xi$-jets passes the two-photon triggers, it is unlikely to survive the event analysis.  Assuming that the isolated photons originate from the Higgs, it is typically required that the invariant mass of the isolated photons is near Higgs mass~\cite{CMS:2021kom, ATLAS:2022vkf}.  However, in the presence of the $\xi$-jets, the isolated photons will fail this requirement.  A re-analysis including the $\xi$-jets would be necessary to reconstruct the Higgs.

\subsection{Detector Efficiencies}

\begin{table}[b]
\begin{center}
\begin{tabular}{|c|c|}
\hline
Channel & ATLAS $p_T$ Requirements  \\
\hline
$1 \gamma$
& $p_{1,T} > 150~\text{GeV}$~\cite{Diehl:2015xeq}
\\
\hline
$2 \gamma$
& $p_{1,T} > 35~\text{GeV}$ and $p_{2,T} > 25~\text{GeV}$~\cite{ATLAS:2022tnm}
\\
\hline
$3 \gamma$
& $p_{1,T} > 15~\text{GeV}$, $p_{2,T} > 15 \text{GeV}$, and $p_{3,T} > 15~\text{GeV}$~\cite{ATLAS:2015rsn}
\\
\hline
$4 \gamma$
& $p_{1,T} > 30~\text{GeV}$, $p_{2,T} > 18~\text{GeV}$, $p_{3,T} > 15~\text{GeV}$, and $p_{4,T} > 15~\text{GeV}$~\cite{ATLAS:2015rsn}
\\
\hline
$5\gamma$
& $p_{i,T} > 15~\text{GeV } (i={1,2,3,4,5})$
\\
\hline
$6 \gamma$
& $p_{i,T} >15~\text{GeV } (i={1,2,3,4,5,6})$ 
\\
\hline
\end{tabular}
\end{center}
\caption{
\label{tab:pTATLAS} 
Transverse momenta thresholds on $\gamma_{\rm iso}$ implemented to estimate ATLAS detector efficiency. Transverse momenta are ordered such that $p_{1,T} > p_{2,T} > p_{3,T} > p_{4,T} > p_{5,T} > p_{6,T}$. The choice for five and six-photon categories is motivated by the photon identification efficiency, which falls to $50 (75)\% $ for $p_T =10 (20)$ GeV  \cite{ATLAS:2018fzd, ATLAS:2019qmc, CMS:2020uim}. }
\end{table}

\begin{table}[b]
\begin{center}
\begin{tabular}{|c|c|}
\hline
Channel & CMS $p_T$ Requirements  \\
\hline
$1 \gamma$
& $p_{1,T} > 145~\text{GeV}$~\cite{CMS:2014rwa}
\\
\hline
$2 \gamma$
& $p_{1,T} > 30~\text{GeV}$ and $p_{2,T} > 18~\text{GeV}$~\cite{CMS:2021kom}
\\
\hline
$3 \gamma$
& $p_{1,T} > 15~\text{GeV}$, $p_{2,T} > 15~\text{GeV}$, and $p_{3,T} > 15~\text{GeV}$~\cite{ATLAS:2015rsn}

\\
\hline
$4 \gamma$
& $p_{1,T} > 30~\text{GeV}$, $p_{2,T} > 18~\text{GeV}$, $p_{3,T} > 15~\text{GeV}$, and $p_{4,T} > 15~\text{GeV}$~\cite{CMS:2022xxa}
\\
\hline
$5\gamma$
& $p_{i,T} > 15~\text{GeV } (i={1,2,3,4,5})$

\\
\hline
$6 \gamma$
& $p_{i,T} >15~\text{GeV } (i={1,2,3,4,5,6})$ 

\\
\hline
\end{tabular}
\end{center}
\caption{
\label{tab:pTCMS}
Transverse momenta thresholds on $\gamma_{\rm iso}$ implemented to estimate CMS detector efficiency. Transverse momenta are ordered such that $p_{1,T} \geq p_{2,T} \geq p_{3,T}\geq  p_{4,T} \geq p_{5,T} \geq p_{6,T}$. The choice for five and six-photon categories is motivated by the photon identification efficiency, which falls to $50 (75) \%$ for $p_T = 10 (20)$ GeV  \cite{ATLAS:2018fzd, ATLAS:2019qmc, CMS:2020uim}.}
\end{table}

We will require that the isolated photons are in the barrel or endcap of the detector, excluding the transition region.  For ATLAS~\cite{ATLAS:2015rsn, Diehl:2015xeq, ATLAS:2022tnm}, this demands
\begin{eqnarray}
|\eta|<1.37\quad{\rm or}\quad 1.52<|\eta|<2.5,\label{eq:ATLASrap}
\end{eqnarray}
and for CMS~\cite{CMS:2014rwa,CMS:2021kom, CMS:2022xxa}
\begin{eqnarray}
|\eta|<1.44\quad{\rm or}\quad 1.57<|\eta|<2.5.\label{eq:CMSrap}
\end{eqnarray}
In addition to requiring photons to be in the barrel or endcap, all isolated photons considered in the analysis are required to leave a minimum energy deposit in the detector.  For CMS, we implement a 2 GeV threshold, which corresponds to the local energy maximum for the ``seed'', trigger tower~\cite{CMS:2020cmk}. For ATLAS, we implement a threshold of 0.5 GeV that corresponds to the local energy maximum of the proto-cluster~\cite{ATLAS:2019qmc}.

The final step is to implement the transverse momentum requirements on $\gamma_{\rm iso}$ of the trigger and analysis. The transverse momentum thresholds used to estimate the ATLAS and CMS detector responses are detailed in Table~\ref{tab:pTATLAS} and Table~\ref{tab:pTCMS}, respectively. If a search has been performed for a given category, we match the transverse momenta requirements of that search. For the three-photon category at CMS, there is no such search, so we use three-photon trigger requirements from the ATLAS search~\cite{ATLAS:2015rsn}. To the author's best knowledge, no five or six-photon search has been performed that requires all photons to pass the trigger. The single photon identification efficiency falls to about $50\%$ when $p_T = 10$ GeV \cite{ATLAS:2018fzd}. When $p_T  = 20$ GeV this efficiency increases to about $75\%$ \cite{ATLAS:2019qmc, CMS:2020uim}. This motivates approximating the five and six-photon triggers by taking $p_T  > 15$ GeV for all isolated photons in the signal.

\begin{figure}[tb]
\begin{center}
    \subfigure[]{\includegraphics[width=0.49\textwidth,clip]{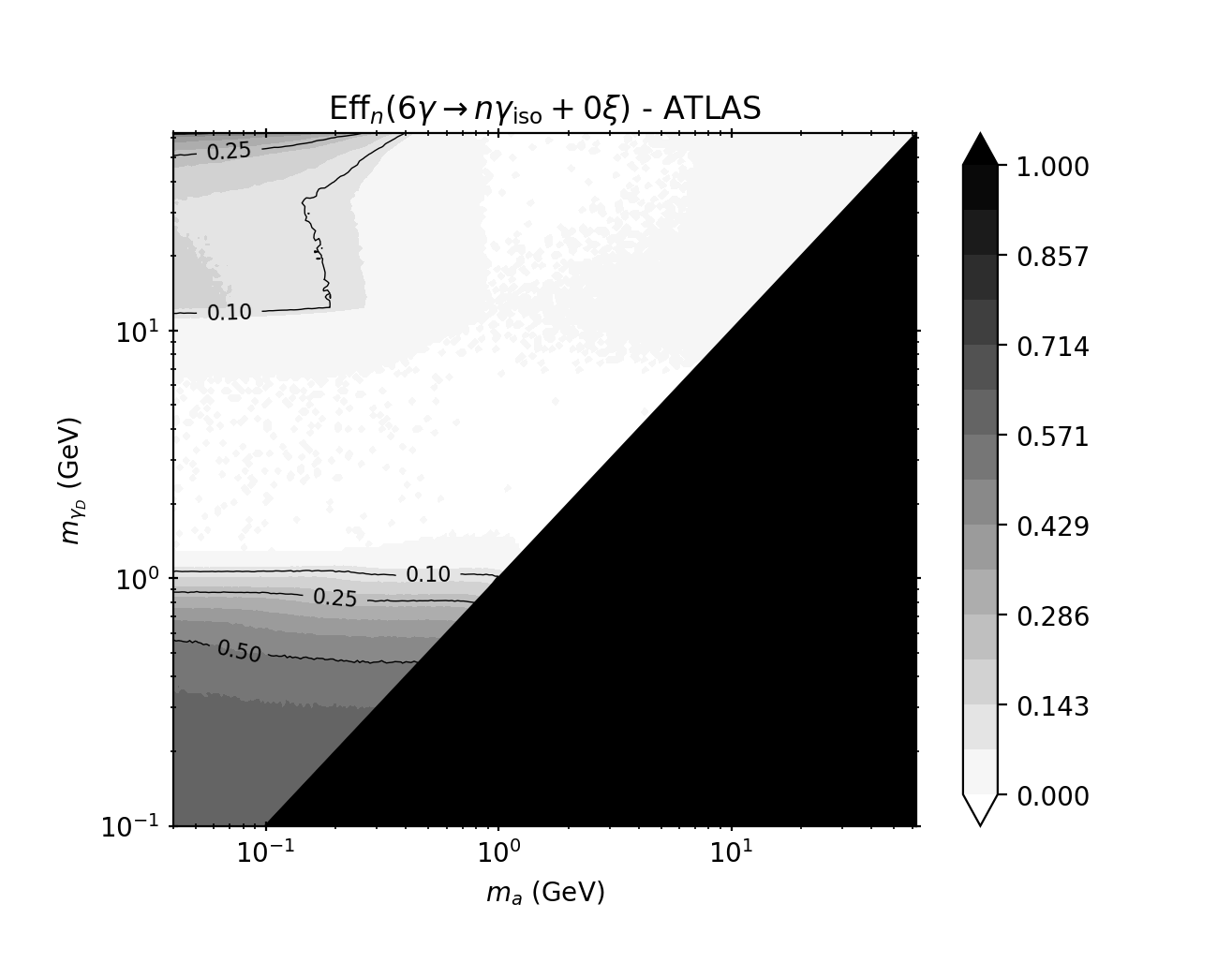}}
    \subfigure[]{\includegraphics[width=0.49\textwidth,clip]{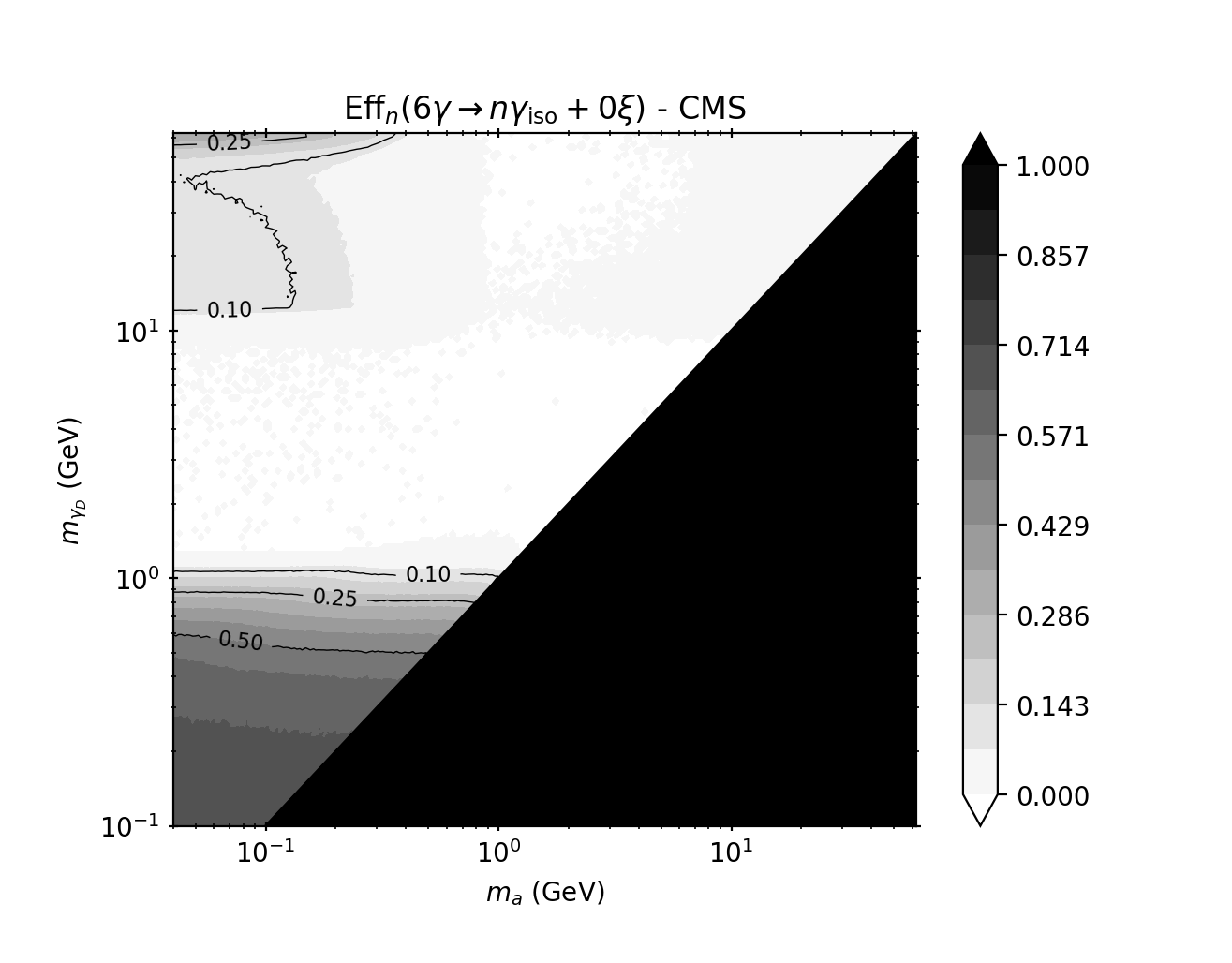}}
\end{center}
\caption{
\label{fig:effEstimate}
Estimated detector efficiencies for $n$ photon triggers applied to $n$ isolated photons and zero $\xi$-jet signals for both (a) ATLAS and (b) CMS. Efficiencies are estimated by placing rapidity [Eqs.~\eqref{eq:ATLASrap}-\eqref{eq:CMSrap}] and transverse momentum requirements on the isolated photons (Tabs.~\ref{tab:pTATLAS}, \ref{tab:pTCMS}).}
\end{figure}

The $k$-photon trigger consists of Eqs.~\eqref{eq:ATLASrap}-\eqref{eq:CMSrap}, minimum $p_T$ requirements on all $\gamma_{\rm iso}$, and the $k$-photon $p_T$ requirements on the hardest photons in Tabs.~\ref{tab:pTATLAS}, ~\ref{tab:pTCMS}.  We define the efficiency Eff$_k(h\rightarrow n\gamma_{\rm iso}+m\xi)$ as the fractions of $n$ isolated photon and $m$ $\xi$-jet events that survive the $k$-photon triggers.   The resulting detector efficiencies for $n$ isolated photons with zero $\xi$-jet for both ATLAS and CMS are shown in Fig.~\ref{fig:effEstimate}. These triggers are implemented such that the $n$-photon trigger is used for signals with exactly $n$ isolated photons. Overall, the two-photon triggers are quite good at finding the two-photon signals in the lower portion of the plot with efficiencies above $10\%$. The four-photon triggers are able to see a comparable number of four-photon signals in the top-left region. We note that for intermediate dark photon mass, $1$ GeV $\alt m_{\gamma_D} \alt 10$ GeV, the detector efficiency is quite bad. The problem is the dominant branching fraction is actually to $0\gamma_{\rm iso} +2\xi$-jets. In this region, there are simply no isolated photons on which to trigger. The efficiencies are also quite bad for the upper right region where $m_a \approx m_{\gamma_D} \approx m_h/2$. This is partially due to the fact that we must reconstruct the Higgs boson mass. This limits the energy available to each individual photon, in particular the soft photons originating from the dark photon decay. Even when the photons from the dark photon are not particularly soft, the Higgs mass is split between the energies of six isolated photons.  If the Higgs mass is split evenly, all six photons have energies $\sim 21$~GeV.  Moving slightly away from an even split forces some photons to become softer.  Hence, requiring $p_T > 15$ GeV for all photons eliminates the majority of the signal.    We will revisit the possibility of triggering on fewer photons than the total number of isolated photons in an event and on events with $\xi$-jets, which could in principle salvage some of the signals.

\subsection{Two-photon Signals}

\begin{figure}
\begin{center}
    \subfigure[]{\includegraphics[width=0.49\textwidth,clip]{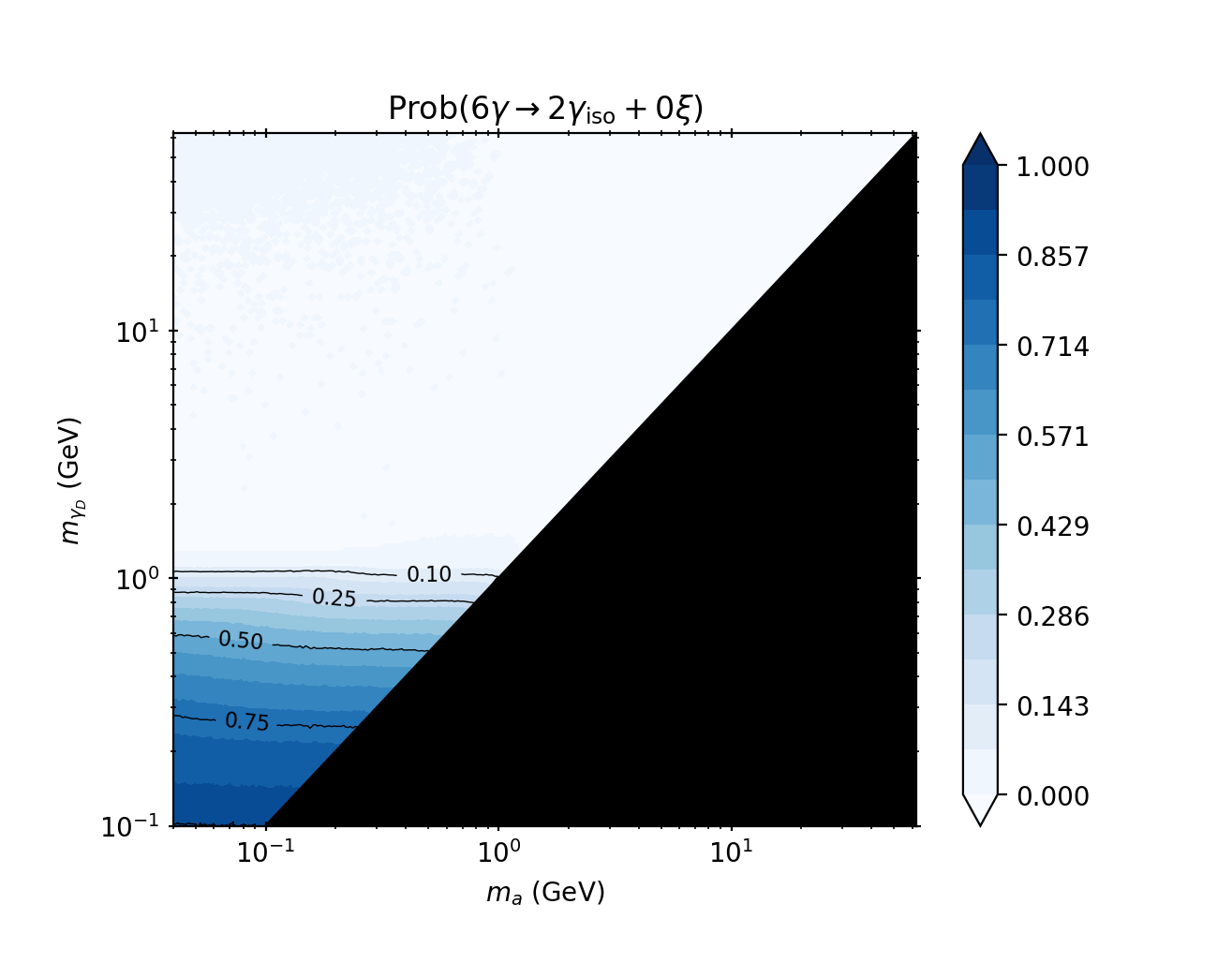}}
    \subfigure[]{\includegraphics[width=0.49\textwidth,clip]{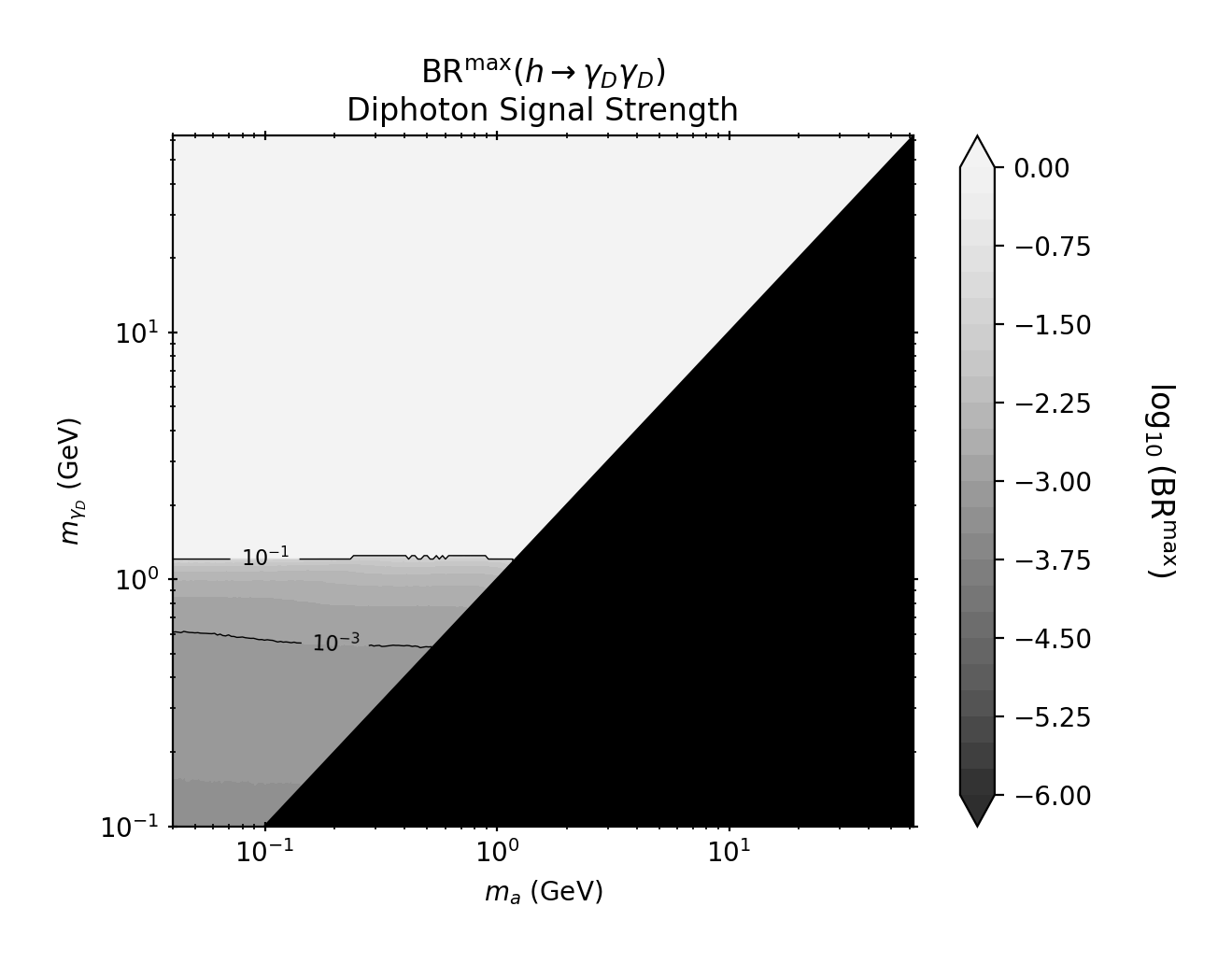}}
\end{center}
\caption{
\label{fig:2phot}
(a) Probability of a signal appearing as a $2\gamma_{\rm iso}+0\xi$-jet final state. (b) Upper bound on the branching fraction BR($h\rightarrow \gamma_D \gamma_D$) from Higgs diphoton signal strength.  A detector efficiency of ${\rm Eff}_2(6\gamma\rightarrow 2\gamma_{\rm iso}+0\xi)\geq 1\%$ is required.   In (b) the contours are upper bounds on ${\rm BR}(h\rightarrow \gamma_D\gamma_D)$ while the heat map is for ${\rm log}_{10}{\rm BR}(h\rightarrow\gamma_D\gamma_D)$.  We assume ${\rm BR}(a\rightarrow \gamma\gamma)={\rm BR}(\gamma_D\rightarrow a\gamma)=1$.
}
\end{figure}

The Higgs to diphoton rate continues to be a cornerstone of Higgs physics. As shown previously, our signal can contribute to the diphoton signal, altering measurements away from the SM prediction.  Measurements of the Higgs to diphoton rate are characterized by the signal strength:
\begin{align}
\mu & = \frac{\sigma(p p \rightarrow h)}{\sigma_\text{SM}(p p \rightarrow h)} 
\frac{\text{BR}(h\rightarrow \gamma \gamma )}{\text{BR}_\text{SM}(h\rightarrow \gamma \gamma)},
\end{align}
where the subscript SM indicates the SM value.  
Measurements restrict the signal strength to be
\begin{align}
\mu = 1.04 \pm ^{0.10}_{0.09} (1.12 \pm ^{0.09}_{0.09})\label{eq:sigstrength}
\end{align}
for ATLAS (CMS) \cite{ATLAS:2022tnm,CMS:2021kom} diphoton rate. 

For our scenario, there will be an additional two-photon signal when $m_{\gamma_D} \alt 1 $ GeV, as shown in Fig.~\ref{fig:2phot}(a). This two-isolated photon signal would appear as an increased Higgs to diphoton partial width.  Taking the production cross section to the SM value, the relevant signal strength for our model is given by

\begin{align}
\mu &= 
\frac{1}
{\text{BR}_\text{SM}(h \rightarrow\gamma \gamma)}
\frac{{\Gamma_\text{SM}(h \rightarrow \gamma \gamma) + \Gamma(h\rightarrow \gamma_D \gamma_D ){\rm BR}^2(\gamma_D\rightarrow a\gamma){\rm BR}^2(a\rightarrow \gamma\gamma) \text{Prob}(6\gamma \rightarrow 2 \gamma_{\rm iso}+0\xi)}
}
{\Gamma_\text{SM}(h) + \Gamma(h\rightarrow \gamma_D \gamma_D)}.
\end{align}
The probability ${\rm Prob}(6\gamma\rightarrow 2\gamma_{\rm iso}+0\xi)$ is defined in Eq.~\eqref{eq:kinBR} and shown in Fig.~\ref{fig:2phot}(a).  Since the dark photon and ALP branching ratios only alter the apparent partial width of Higgs into diphotons but not the Higgs total width, we will assume ${\rm BR}(\gamma_D\rightarrow a\gamma)={\rm BR}(a\rightarrow\gamma\gamma)=1$ for simplicity.  As shown previously, the dark photon and ALP are still expected to decay before the ECAL in the parameter region under consideration.  It should also be noted that the limit on the Higgs to the unknown branching fraction does not apply for our $2\gamma_{\rm iso}+0\xi$-jet signal. This is because this signal modifies the apparent Higgs to diphoton partial width and not just the total width.  Hence, the $2\gamma_{\rm iso}+0\xi$-jet signal alters the diphoton signal strength differently than other Higgs signal strengths.  We estimate the region of validity for the unknown branching fraction limit as the region where less than $1\%$ of two-isolated photon signals pass the two-photon triggers.    Nevertheless, we will show that the diphoton measurement is more constraining than the Higgs to the unknown limit.

We fit to the signal strengths in Eq.~(\ref{eq:sigstrength}) and place 95\% CL bounds on ${\rm BR}(h\rightarrow \gamma_D\gamma_D)$.  These results are shown in Fig.~\ref{fig:2phot}(b).
In obtaining these limits, we required that the detector efficiency of the two-photon trigger be larger than 1\% for our two-isolated photon and zero-$\xi$-jet signal.  The limits are quite good.  The branching ratio is limited to be less than $10^{-3}$ over much of parameter space where the $2\gamma_{\rm iso}+0\xi$-jet signal is dominant.   The branching ratio limits weaken significantly as the other signal classes become more relevant.  However, the branching ratio limits from Higgs to diphoton measurements are still better than the Higgs to the unknown upper limit of 0.12.

\subsection{Four-photon Signals}
\begin{figure}
\begin{center}
    \subfigure[]{\includegraphics[width=0.49\textwidth,clip]{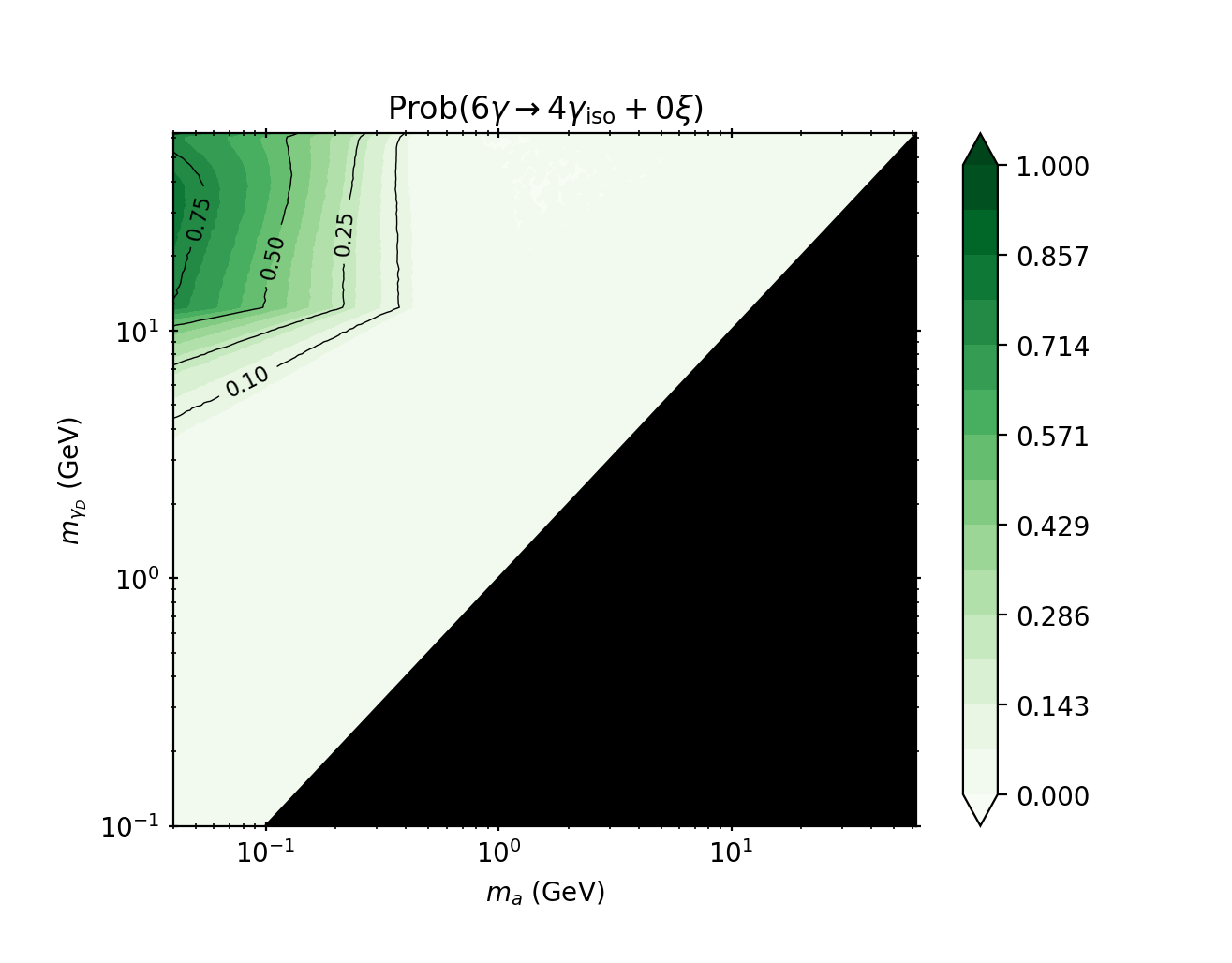}}
    \subfigure[]{\includegraphics[width=0.49\textwidth,clip]{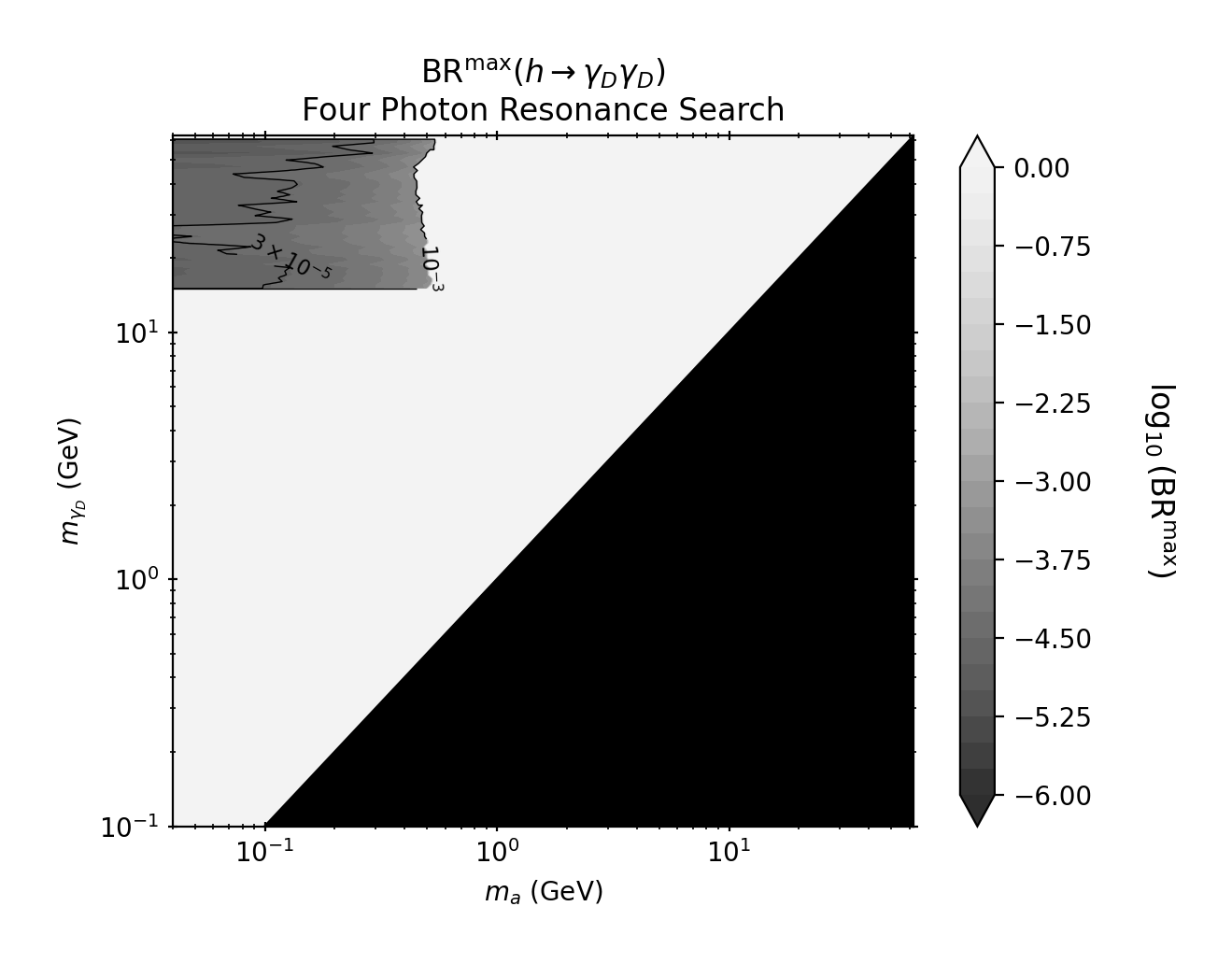}}
\end{center}
\caption{
\label{fig:4phot}
(a) Probability of four-isolated photon + zero-$\xi$-jet final state. (b)  Constraint (dark area) on the branching fraction BR($h\rightarrow \gamma_D \gamma_D$) using the Higgs four-photon resonant search from CMS \cite{CMS:2022xxa} and requiring $1\%$ detector efficiency for the CMS four-photon trigger. In (b) the contours are upper bounds on ${\rm BR}(h\rightarrow \gamma_D\gamma_D)$ while the heat map is for ${\rm log}_{10}{\rm BR}(h\rightarrow\gamma_D\gamma_D)$.  We assume ${\rm BR}(a\rightarrow \gamma\gamma)={\rm BR}(\gamma_D\rightarrow a\gamma)=1$.
}
\end{figure}

The Higgs to four-photon decay is very rare in the SM.  The four-photon branching fraction is suppressed by $\sim(\alpha_\text{EM}/4\pi)^2$ relative to the diphoton rate. The major backgrounds in these signals are the SM production of $\gamma \gamma +$jets, $\gamma +$jets, as well as multi-jet events, in which jets are misidentified as isolated photons. As such, even a small observed rate could be evidence of the BSM physics.

In our scenario, four-photon signals are prevalent when $m_{\gamma_D} \agt 10$ GeV and $m_a \alt 100$ MeV. This can be seen in Fig.~\ref{fig:4phot}(a) which shows the probability that our six-photon signal results in four isolated photons and zero $\xi$-jet. These events occur when the photons originating from the ALP decay become extremely boosted, forming a photon-jet. Hence, the dark photon decays into a photon and a photon-jet.  As discussed previously, this would be an apparent violation of the Landau-Yang theorem.  

We take the constraints from a recent CMS search for the Higgs decaying into a pair of ALPs which subsequently decay into photons: $h\rightarrow aa\rightarrow 4\gamma$~\cite{CMS:2022xxa}.   
This search is relevant for $15~{\rm GeV}\leq m_{\gamma_D}\leq 62~{\rm GeV}$.  The results are reported as a limit on the Higgs cross section times branching ratio.  For the Higgs production cross section, we use the inclusive rate in Eq.~(\ref{eq:xsect}).  The four-photon branching fraction in our model is given by 
\begin{align}
\text{BR}(h\rightarrow 4 \gamma) = \text{BR}(h\rightarrow \gamma_D \gamma_D){\rm BR}^2(\gamma_D\rightarrow a\,\gamma){\rm BR}^2(a\rightarrow \gamma\gamma) \text{Prob}(6\gamma \rightarrow 4 \gamma_{\rm iso}+0\xi).
\end{align}
As before with the two-photon signal, we will assume ${\rm BR}(\gamma_D\rightarrow a\gamma)={\rm BR}(a\rightarrow \gamma\gamma)=1$.  We require the detector efficiency for the CMS four-photon trigger to be larger than 1\% [Fig.~\ref{fig:effEstimate}(b)] to guarantee sufficient statistics.  However, since the reported experimental limits are unfolded, we do not apply the detector efficiencies to our results.

The recast limits on ${\rm BR}(h\rightarrow \gamma_D\gamma_D)$ are shown in Fig.~\ref{fig:4phot}(b). The drop-off in sensitivity at the top of the plot is due to the experimental search stopping at $62$ GeV, while our plots go to $m_h/2 = 62.5$ GeV.  The constraint from the four-photon resonance search restricts the branching ratios to be lower than $10^{-3}$ over the large majority of the relevant mass range. Constraints get considerably stronger as the probability of a $4\gamma_{\rm iso}+0\xi$-jet signal increases, as shown by the interior contour of $3 \times 10^{-5}$. These are much better than $95\%$ CL bound on Higgs to the unknown fraction of $0.12$. 

\subsection{Six-photon Signals}
\begin{figure}[tb]
\begin{center}
    \subfigure[]{\includegraphics[width=0.49\textwidth,clip]{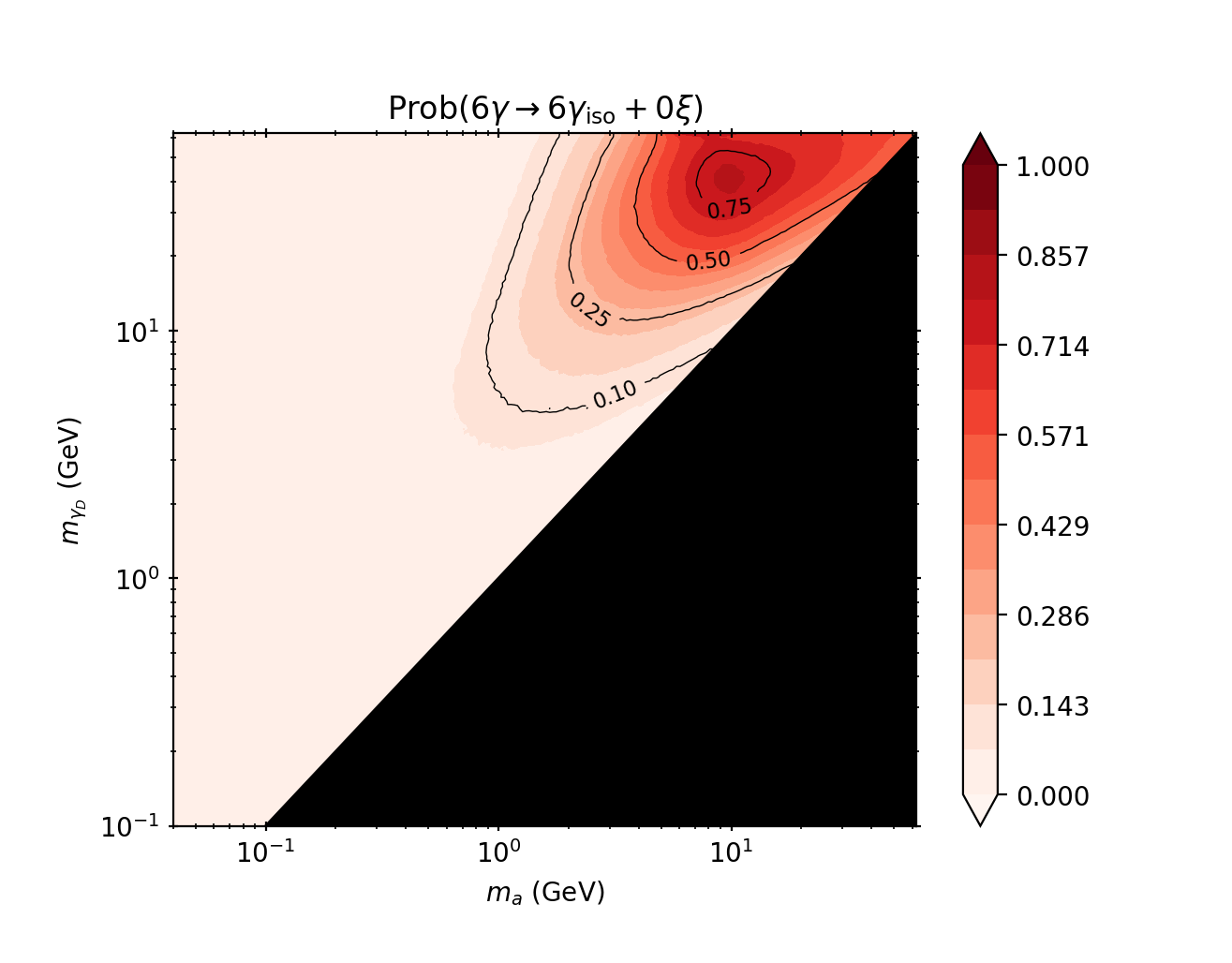}}
    \subfigure[]{\includegraphics[width=0.49\textwidth,clip]{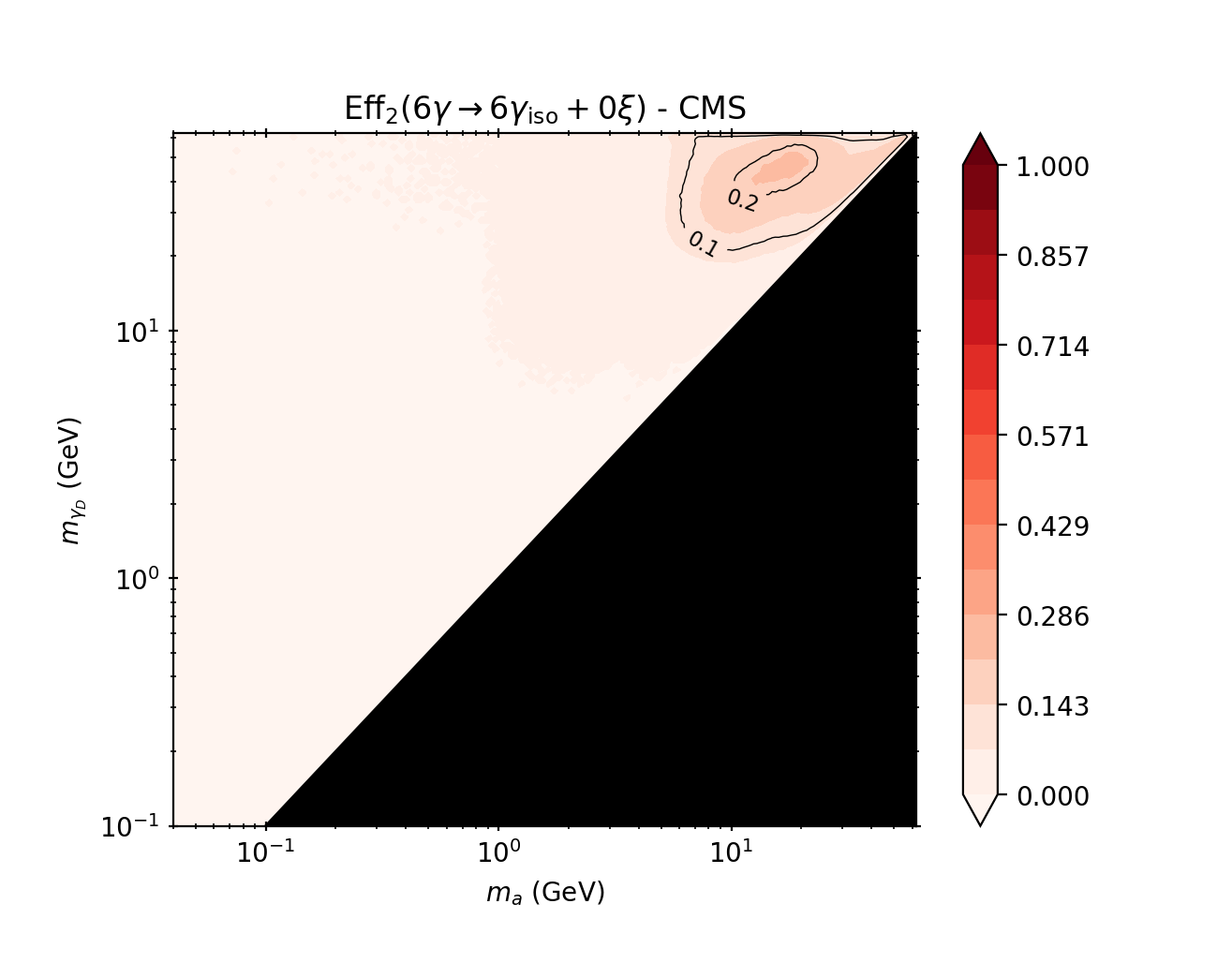}}
\end{center}
\caption{
\label{fig:6phot}
(a) Probability of signals becoming a six-isolated photons + zero-$\xi$-jet final state. (b) Estimated CMS detector efficiency for six-photon signals with two-photon triggers.}
\end{figure}

While the SM four-photon Higgs resonance is exceptionally rare, the six-photon resonance is even rarer. The SM branching fraction is further suppressed by roughly another factor of $\sim(\alpha_\text{EW} / 4\pi)^2$ with respect to the four-photon rate.  The dominant backgrounds will be from the SM production of multi-photon, multi-photon plus multi-jet, and multi-jet, where jets fake photons. 

In our scenario, six-photon signals dominate when $m_{\gamma_D} \agt 10$ GeV and $m_a \agt 1$ GeV. This can be seen in Fig.~\ref{fig:6phot}(a). In this regime, all photons remain well-separated. To the authors' best knowledge, a six-photon Higgs resonance search has not been performed. However, one can still place constraints using the Higgs fits.   In particular, the bound on Higgs decaying into undetermined final states is relevant for our six-photon signal: ${\rm BR}(h\rightarrow \gamma_D\gamma_D)\leq 0.12$ where, as before, we assume ${\rm BR}(a\rightarrow \gamma\gamma)={\rm BR}(\gamma_D\rightarrow a\gamma)=1$.

The Higgs to six-isolated photon signal motivates new searches at the LHC.  However, as can be seen in the top-right regions of Fig.~\ref{fig:effEstimate}(a) and (b), the detector efficiency for imposing a six-photon trigger is quite small. It is at best a few percent, as discussed previously. To improve detector efficiency, we investigated whether the acceptance could be improved by lowering the $p_T$ requirements of the softest photons to $5$~GeV: $p_{5,T},p_{6,T}\geq 5$~GeV. However, this only offers improvement to about $5\%$ for $m_a \approx m_{\gamma_D} \approx m_h /2$.  In principle, the transverse momentum thresholds on all photons could be lowered. However, as discussed above, the detector's ability to identify photons falls off quickly at low $p_T$ \cite{CMS:2021kom,ATLAS:2018fzd,ATLAS:2019qmc}, making the background suppression significantly more difficult. 

One of the most promising avenues to retain the six-photon signal is to loosen the number of photons to be triggered on.  For example, using the two-photon trigger significantly improves the efficiency to  $\gtrsim 10\%$ over a good portion of the six-photon signal region. Such a trigger scenario is shown in Fig.~\ref{fig:6phot}(b). Some of these events might have been saved from the Higgs to diphoton searches.  However, since the two hardest photons do not reconstruct the Higgs mass, it is unlikely these events survived further analysis beyond the triggers.  One could try to reanalyze any saved events that passed the two-photon trigger, looking for six photons that reconstruct the Higgs mass.

\subsection{$\xi-$Jet Signals}

\begin{figure}
\begin{center}
    \subfigure[]{\includegraphics[width=0.49\textwidth,clip]{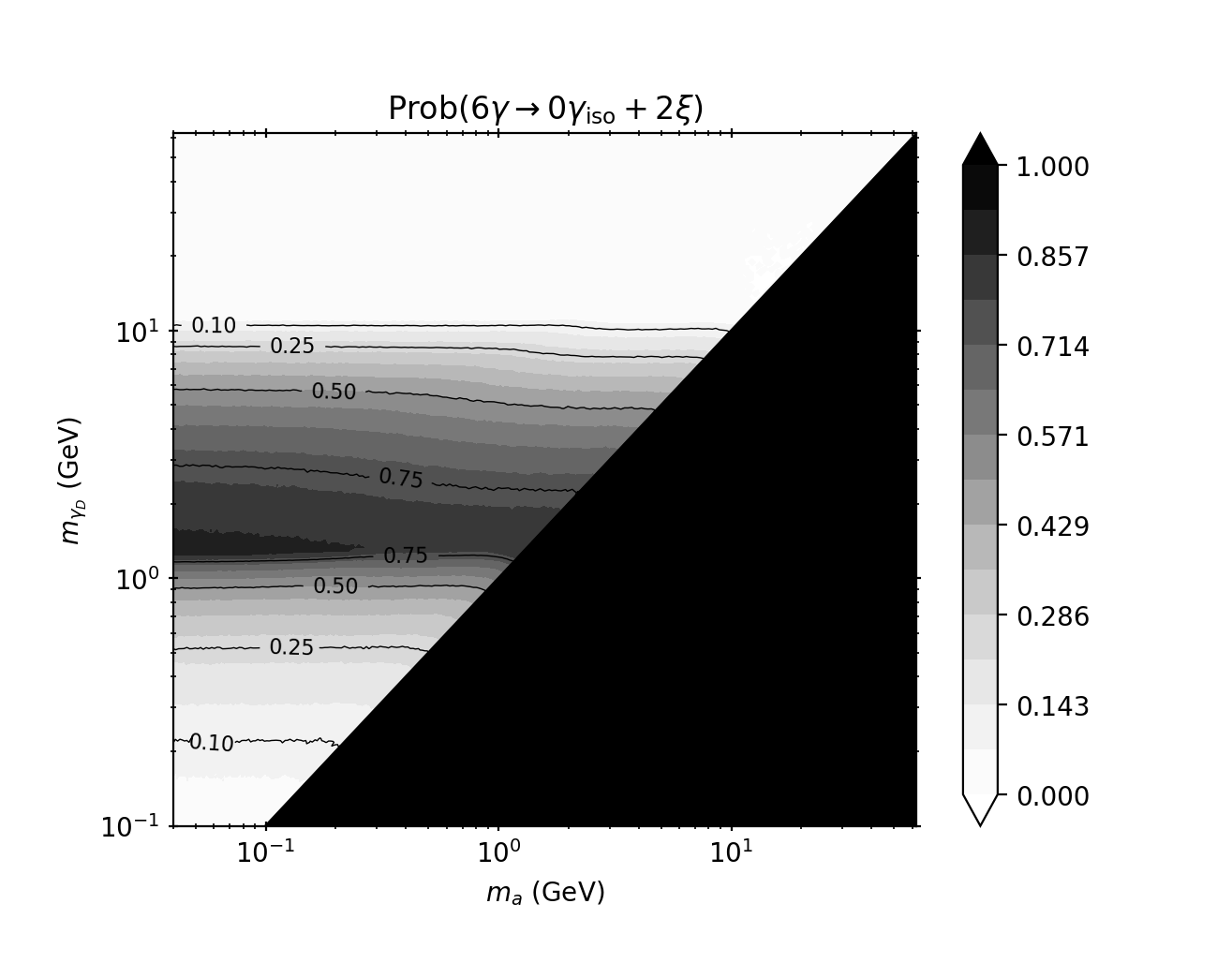}}
    \subfigure[]{\includegraphics[width=0.49\textwidth,clip]{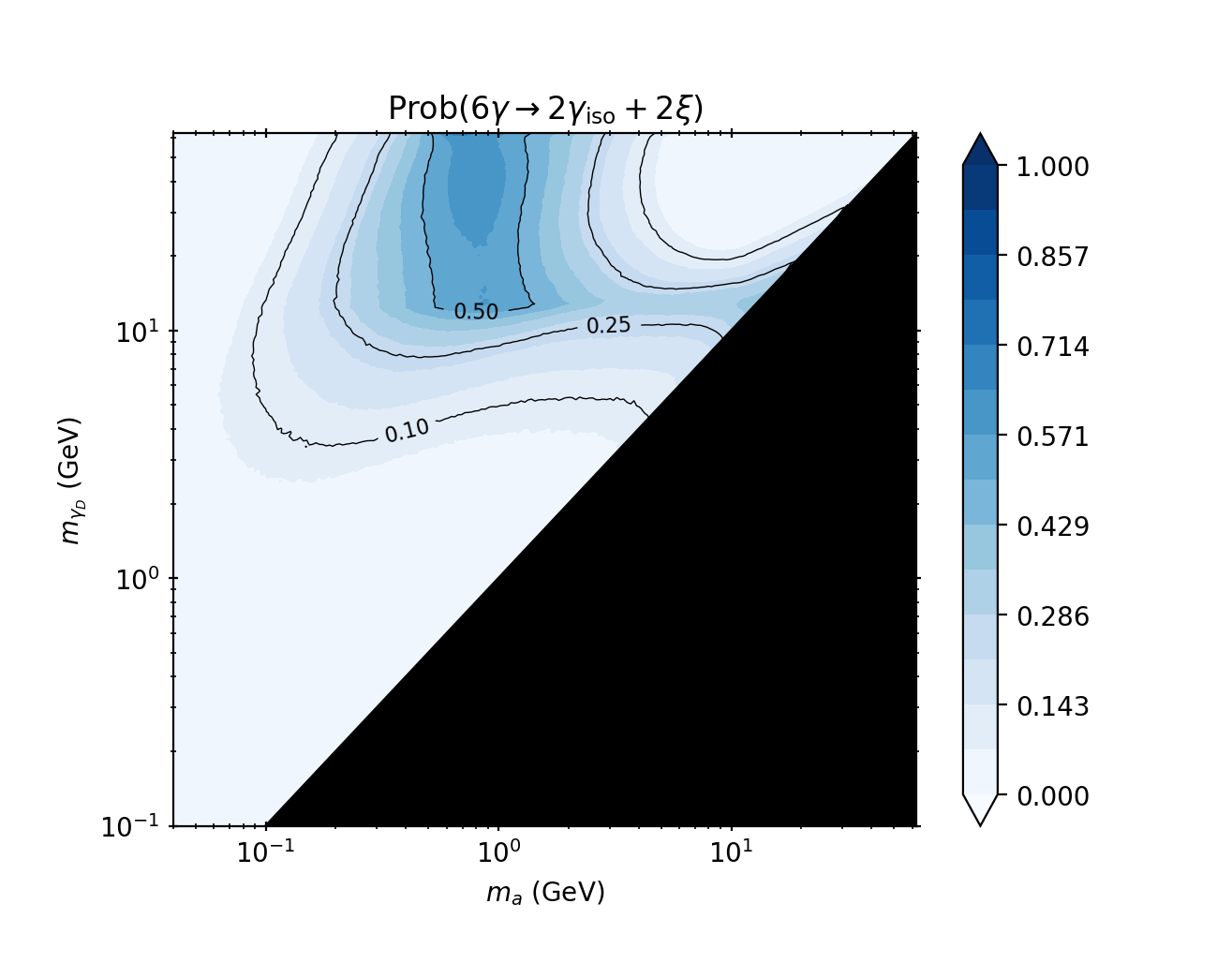}}
    \\
    \subfigure[]{\includegraphics[width=0.49\textwidth,clip]{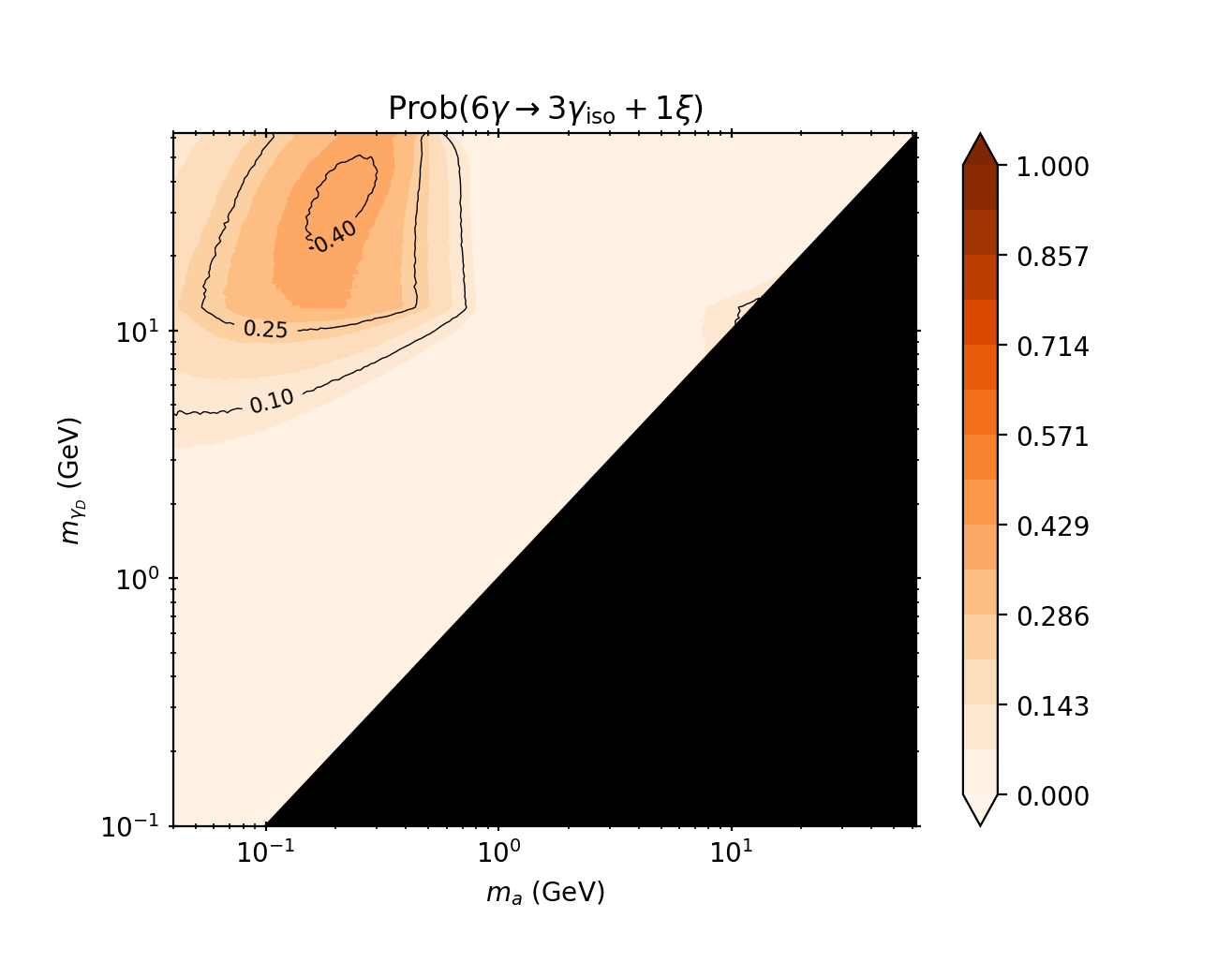}}
    \subfigure[]{\includegraphics[width=0.49\textwidth,clip]{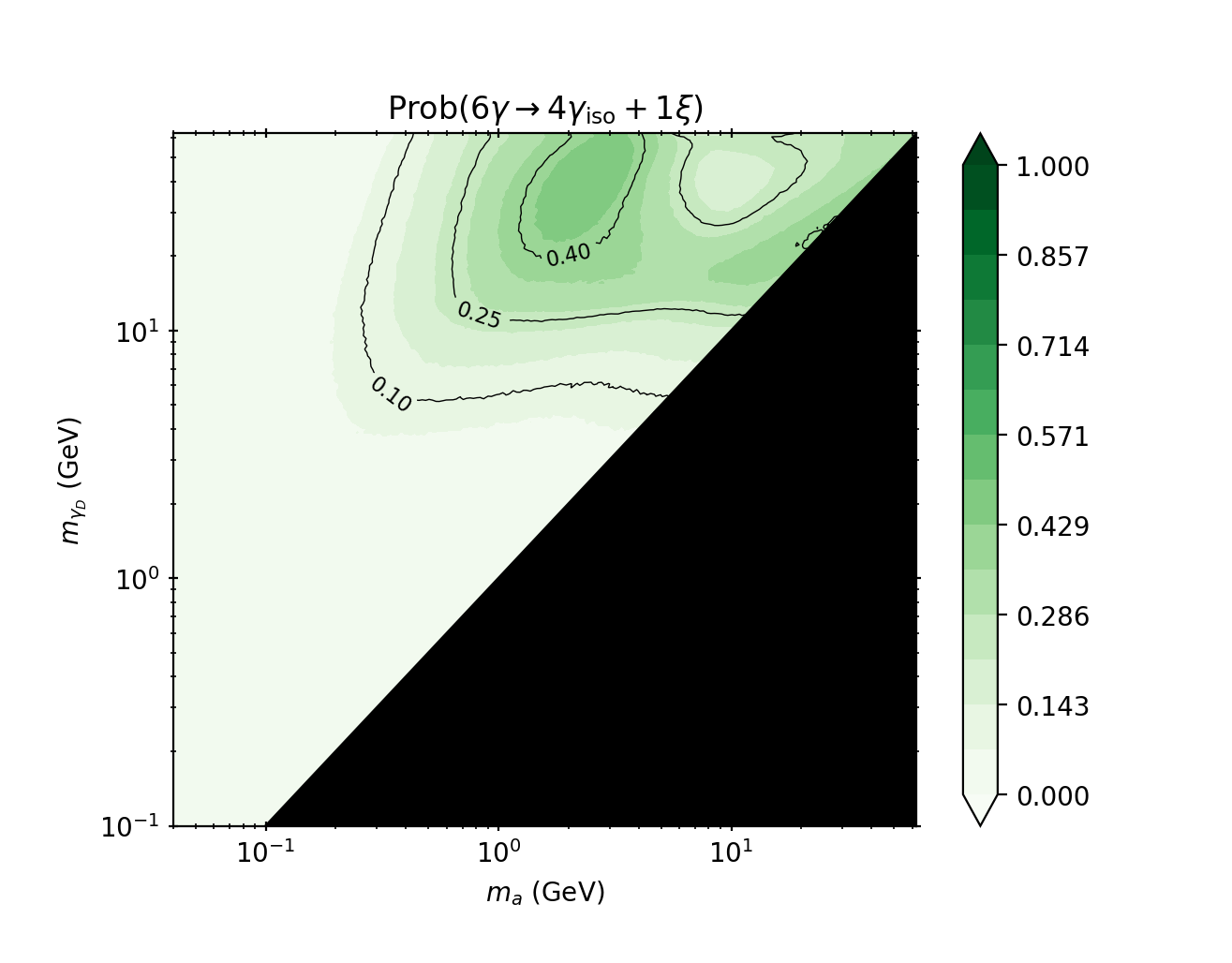}}
\end{center}
\caption{
\label{fig:otherBR}
Probabilities of signals with (a) zero isolated photon plus two $\xi$-jets final state, (b) two isolated photons plus two $\xi$-jets final state, (c) three isolated photons plus one $\xi$-jet final state, and (d) four isolated photons plus one $\xi$-jet final states. }
\end{figure}

We have already seen many interesting multi-photon resonances in our scenario. We would also like to highlight some of the more challenging signals that contain intermediately separated photons, i.e. $\xi$-jets. Namely, the zero-photon plus two $\xi$-jets, the three-photon plus one $\xi$-jet, the four-photon plus one $\xi$-jet, and the two-photon plus two $\xi$-jets signals. The non-negligible branching ratios for these signals are shown in Fig.~\ref{fig:otherBR}. There are certain regions in the $m_a$-$m_{\gamma_D}$ plane for which these types of signals are dominant. Indeed, it is these signals which mainly account for the gray and dark gray regions in Fig.~\ref{fig:DomBRIso}. The $\xi$-jets themselves do not meet the normal isolation criterion for photons.  Despite this, we will show that some of these signals could very well have been saved to tape. However, since they are not identified as photon candidates, the analysis might not reconstruct the correct Higgs mass even if they passed the triggers.  Hence, they may not survive the standard analysis. If this is the case, current data could be reanalyzed to search for additional signals of new physics.

To find the zero-photon and two $\xi$-jets signals, one would need to be able to reconstruct $\xi$-jets or at least trigger on them.  Some recent work has investigated $\xi$-jet triggers and identification \cite{Bowen:2007ia}. Without a full detector simulation, we will ignore pure $\xi$-jet signals. For three-photon plus one $\xi$-jet, four-photon plus one $\xi$-jet, and the two-photon plus two $\xi$-jets signals, we investigate whether the isolated photons could pass the triggers. 

\begin{figure}
\begin{center}
    \subfigure[]{\includegraphics[width=0.49\textwidth,clip]{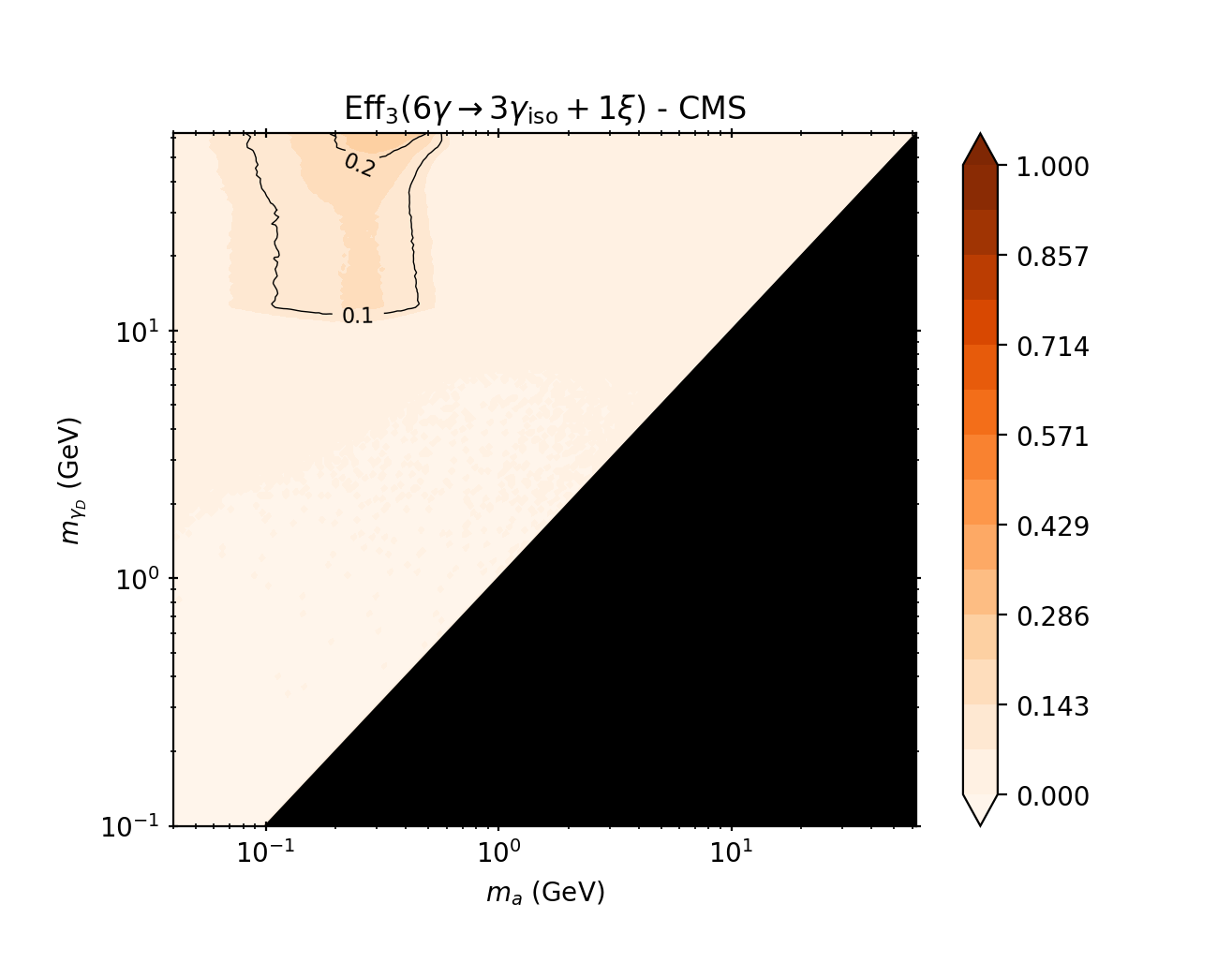}}
    \subfigure[]{\includegraphics[width=0.49\textwidth,clip]{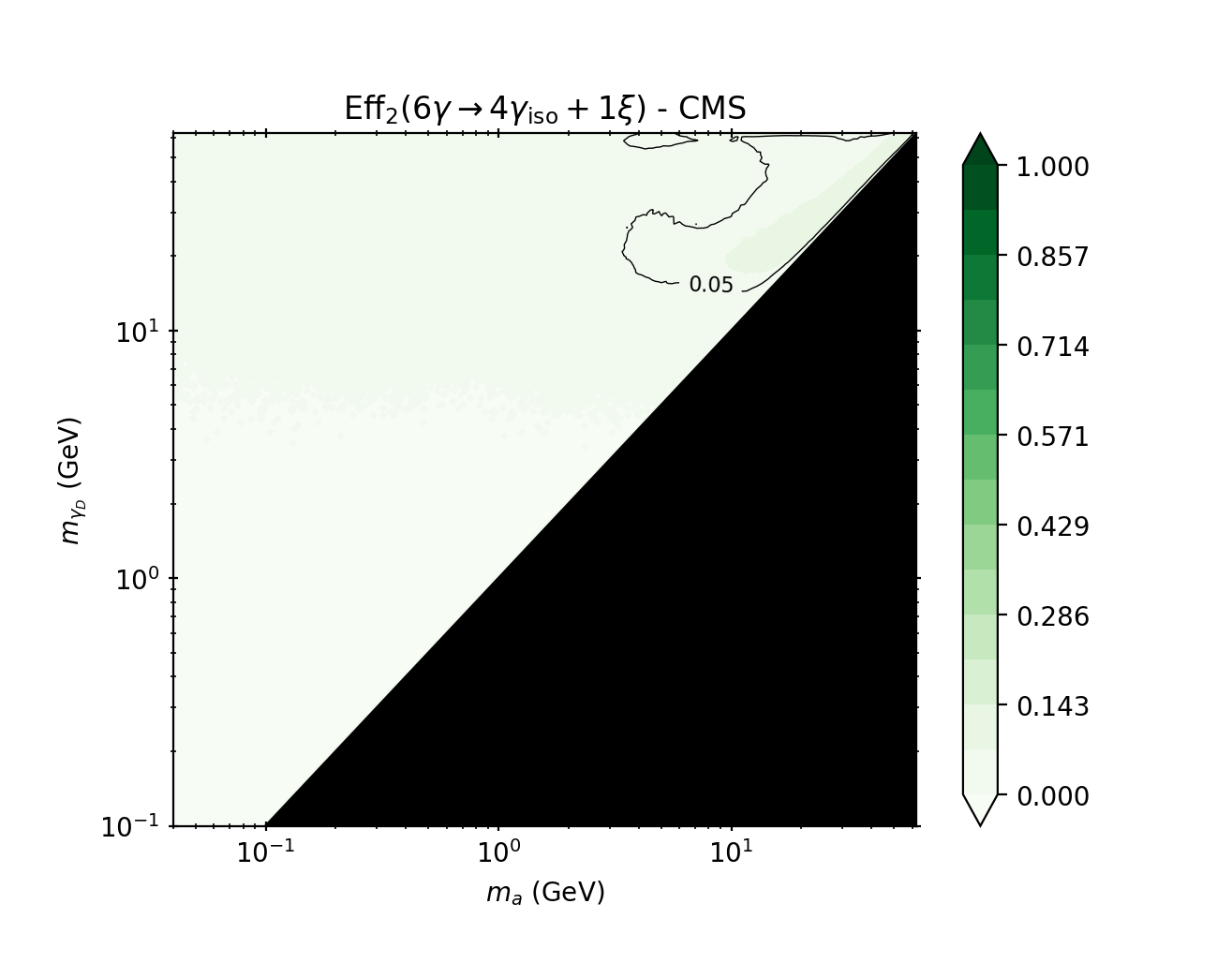}}
    \subfigure[]{\includegraphics[width=0.49\textwidth,clip]{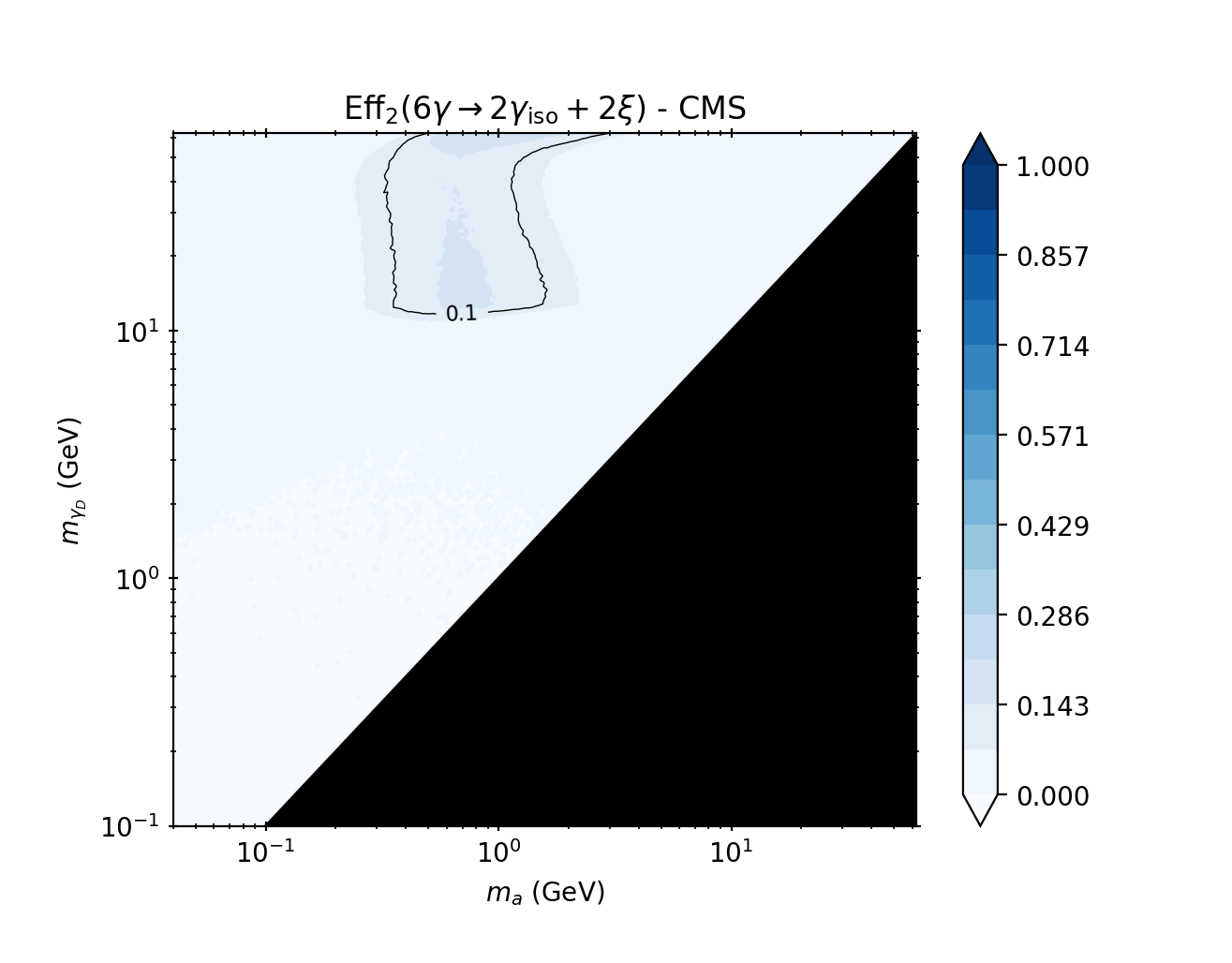}}
\end{center}
\caption{
\label{fig:otherEff}
Estimated CMS detector efficiency for (a) three-photon and one $\xi$-jet signal with the three-photon trigger, (b) four-photon and one $\xi$-jet signal with the two-photon trigger, and (c) two-photon and two $\xi$-jets signal with the two-photon trigger.}
\end{figure}

In Fig.~\ref{fig:otherEff}, we show the efficiencies of (a) the $3\gamma_{\rm iso}+1\xi$-jet passing the CMS three-photon trigger, (b) the $4\gamma_{\rm iso}+1\xi$-jet signal passing the CMS two-photon trigger, and (c) the $2\gamma_{\rm iso}+2\xi$-jet signal passing the CMS two-photon trigger.  In the case of the four-photon plus one $\xi$-jet, we found that almost no events passed the three-photon or four-photon triggers. As such, we focused on the two-photon triggers for this category. 
For all these categories, some events survive the estimated CMS trigger. For the three-photon plus one $\xi$-jet category, about $10\%$ of events survive in the mass window $m_{\gamma_D} \agt 10$ GeV and $0.08$ GeV $\alt m_a \alt 0.6 $ GeV. For four-photon plus one $\xi$-jet, only about $5\%$ of events survive when $m_{\gamma_D} \agt 10$ GeV and $m_a \agt 4 $ GeV. Finally, for two-photon plus two $\xi$-jets, about $10\%$ of events survive when $m_{\gamma_D} \agt 10$ GeV and $0.3$ GeV $\alt m_a \alt 2 $ GeV.  Even though this efficiency may be relatively small, as shown in the two and four-photon analyses, dedicated searches in this efficiency range can place very stringent limits on ${\rm BR}(h\rightarrow \gamma_D\gamma_D)$.

For any of these signals in their appropriate mass range, some events might have been saved on tape.  However, similarly to the six-photon signal, such events would likely not survive the standard analysis, since the $\xi$-jet's four-momentum is needed to reconstruct the Higgs mass. In this case, they present a prime candidate for a new analysis of existing data. There could be new physics hiding in the data we have already collected.

\section{Conclusion}
\label{sec:conc}
Many BSM scenarios consider new light particles that both couple to the SM Higgs and decay primarily to photons. Any such model will introduce additional multi-photon Higgs resonances. 
This paper has highlighted the multi-photon Higgs resonances that rise from the combination of the Higgs portal (Higgs-dark Higgs mixing), the axion portal (ALP-photon-photon vertex), and the dark axion portal (ALP-photon-dark photon vertex).  These portals generically appear in Higgsed dark-photon extensions of the SM with portal matter charged under the SM and new $U(1)$ gauge groups.  
With these portals, many multi-photon resonances can be produced via the six-photon decay chain of the SM Higgs boson
\begin{eqnarray}
h\rightarrow \gamma_D\gamma_D\rightarrow \gamma a\gamma a\rightarrow 6\gamma.
\end{eqnarray}
Depending on the mass ratios, this signal can give rise to photons sufficiently collimated to be indistinguishable in the detector (photon-jets) and intermediately separated photons that cannot be isolated ($\xi$-jets).  We classified the signals according to the number of isolated photons (photons and photon-jets) and $\xi$-jets.  Of particular interest is when the ALP is sufficiently boosted such that its decay products appear to be a single photon. This gives rise to a four-photon signal, with the decay of a dark photon into a photon and a photon-jet.  Such a decay would be an apparent violation of the Landau-Yang theorem.

\begin{figure}
\begin{center}
    \subfigure[]{\includegraphics[width=0.49\textwidth,clip]{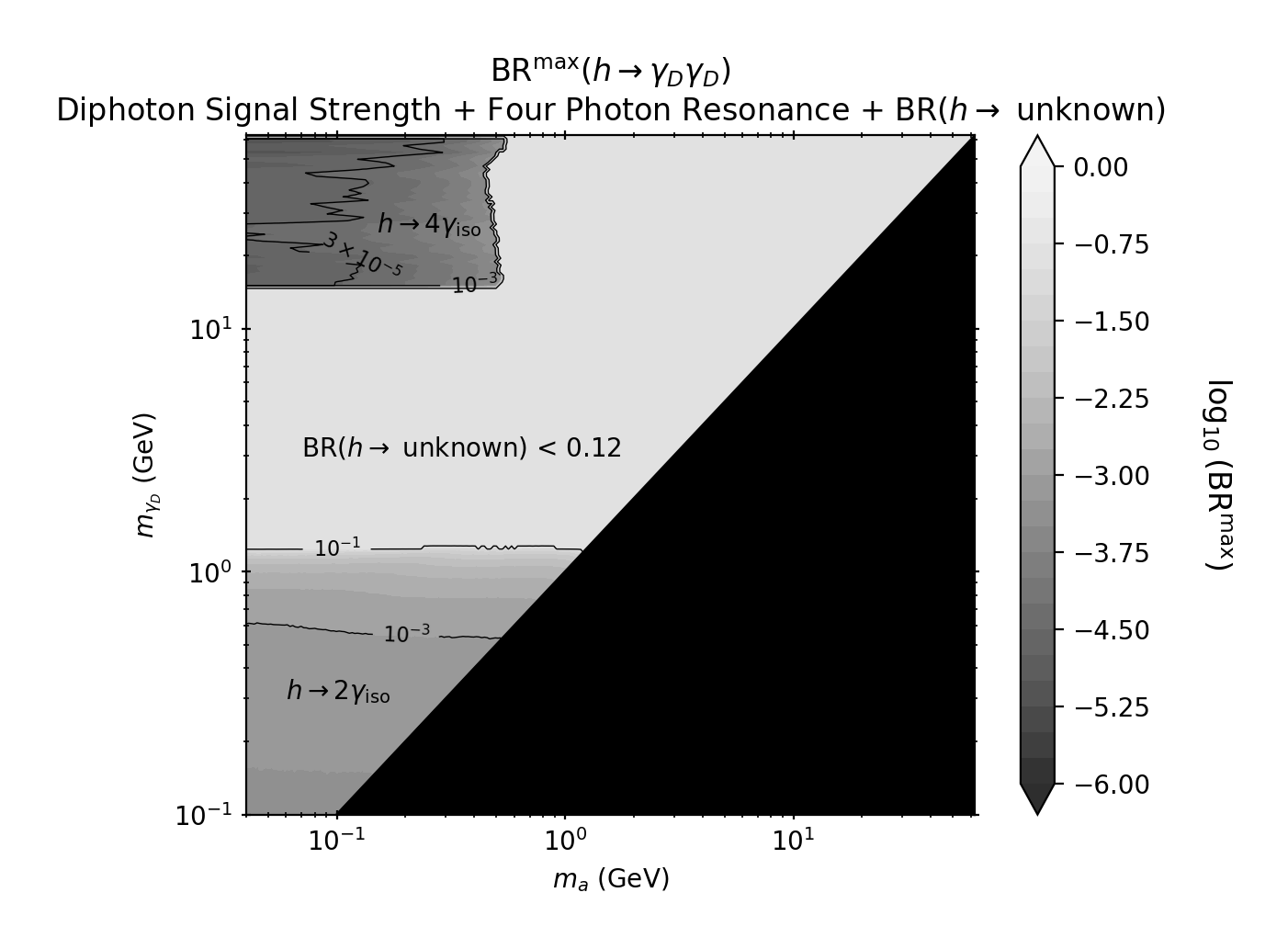}}
\end{center}
\caption{
\label{fig:allConst}
Best constraint on the branching fraction BR($h\rightarrow \gamma_D \gamma_D$) using the Higgs diphoton resonance search, four-photon resonant search, Higgs mixing angle limits, and Higgs branching ratio to the unknown. The resonance searches require $1\% $ detector efficiency. The contours are upper bounds on ${\rm BR}(h\rightarrow \gamma_D\gamma_D)$ while the heat map is for ${\rm log}_{10}{\rm BR}(h\rightarrow\gamma_D\gamma_D)$.  We assume ${\rm BR}(a\rightarrow \gamma\gamma)={\rm BR}(\gamma_D\rightarrow a\gamma)=1$. 
} 
\end{figure}

Using current experimental results, we placed constraints on the branching ratio BR($h\rightarrow \gamma_D \gamma_D$), while assuming BR($\gamma_D\rightarrow a \gamma$) $=$ BR($a\rightarrow \gamma \gamma$) $= 1$. We recast measurements of Higgs to diphoton and searches for Higgs to four photons.  While there are no dedicated searches for Higgs to six isolated photons or with final state $\xi$-jets, we can still limit the branching ratio using the limits to the Higgs to the unknown rate.  The combined limits on our signal in Fig.~\ref{fig:allConst}. 
The center-left, center, and upper-right sections of the kinematically allowed regions are dominated by six-isolated photon or $\xi$-jet signals.  Hence, the upper bound on the $h\rightarrow \gamma_D\gamma_D$ branching ratio comes from the Higgs to the unknown upper limit of $0.12$. In the bottom region, there are observable two-photon signals. Here, Higgs diphoton measurements are able to constrain the branching ratio to $10^{-3}$ when  $m_{\gamma_D} \alt 0.6 $ GeV. The upper-left constraints come from the Higgs to four-photon search. The four-photon resonant searches are able to constrain the branching ratio to about $10^{-3}$ over this entire region. This constraint improves to $3\times 10^{-5}$ when $m_a \lesssim 0.1$ GeV. 

For much of our parameter space, there are interesting signals with six-isolated photons or intermediately separated $\xi$-jets.  These signals are not covered by current direct searches, motivating new searches for exotic Higgs decays.  The six-isolated photon signal provides a particular challenge in that the photons often do not have enough energy to pass a six-photon trigger.  Despite these difficulties, we have shown many of the six-isolated photon and $\xi$-jet events can pass two or three-photon triggers.  Hence, this signal may already be in data that has been recorded.  However, since the two or three hardest photons do not reconstruct the Higgs mass, the six-isolated photon and $\xi$-jet signals are unlikely to survive the standard analyses.  In these scenarios, it could be possible to find new physics by reanalyzing current data. 

The combination of the axion, dark axion, and Higgs portals has provided many interesting and novel signals containing photons at the Higgs pole. Such signals provide challenges in search strategies.  We have recast current searches/measurements to place limits on the model, and have motivated new searches for exotic Higgs decays. One could hope to find evidence of new physics in new searches for multi-photon Higgs resonances, or even by reconsidering existing data.

\section*{Acknowledgements}
IML would like to thank KAIST and the Pittsburgh Particle Physics Astrophysics and Cosmology center for their hospitality during the completion of this paper.  SDL and HL are supported in part by the National Research Foundation of Korea (Grant No. NRF-2021R1A2C2009718).  IML is supported in part by the U.S. Department of Energy under grant No. DE-SC0017988.  The data to reproduce the plots has been uploaded with the arXiv
submission or is available upon request.

\appendix
\section{Branching Ratio Details}
\label{sec:BR}\subsection{Matrix Elements}
\subsubsection{
$ h(\hat{s}) \rightarrow \gamma_D(k_1) ~ \gamma_D(k_2) \rightarrow \big( ~ a(q_1) ~ \gamma(p_1) ~ \big) ~ \big( ~ a(q_2) ~ \gamma(p_2) ~ \big) $
}
Using the Feynman gauge for the propagators, the matrix element for a Higgs of energy $\sqrt{\hat{s}}$ going to two ALPs and two photons is given by
\begin{align}
\nonumber
i\mathcal{M} & = \frac{1}{4} 
\big( i \lambda_{h\gamma_D\gamma_D} g_{\alpha_1 \beta_1} \big)
\left( -i \frac{g^{\alpha_1 \alpha_2} }{k_1^2 - m_{\gamma_D}^2 + i m_{\gamma_D} \Gamma_{\rm total}(\gamma_D)} \right)
\left( -i \frac{g^{\beta_1 \beta_2} }{k_2^2 - m_{\gamma_D}^2 + i m_{\gamma_D} \Gamma_{\rm total}(\gamma_D) } \right)
\\ & \ \ \ \ \ \ \ \ \ \ \ \ \  \ \ \ \times
\left( -i G_{a\gamma\gamma_D} 
\epsilon_{\alpha_2 \mu \rho_1 \sigma_1} p_1^{\rho_1} k_1^{\sigma_1}  \right)
\epsilon_{\mu}^{*}(\sigma_1, p_1)
\\ & \ \ \ \ \ \ \ \ \ \ \ \ \ \ \ \ \times
\left( -i G_{a\gamma\gamma_D} \epsilon_{\beta_2 \nu \rho_2 \sigma_2} p_2^{\rho_2} k_2^{\sigma_2}  \right)
\epsilon_{\nu}^{*}(\sigma_2, p_2) .
\nonumber
\end{align}
Squaring and averaging over the polarization,
\begin{align}
|\mathcal{\overline{M}}|^2 & = 
\frac{|\lambda_{h\gamma_D\gamma_D}|^2 |G_{a\gamma\gamma_D}|^4 }{64 
\left( (k_1^2 - m_{\gamma_D}^2)^2 + m^2_{\gamma_D} \Gamma^2_{\rm total}(\gamma_D)  \right) 
\left( (k_2^2 - m_{\gamma_D}^2)^2 + m^2_{\gamma_D} \Gamma^2_{\rm total}(\gamma_D)  \right)} 
\nonumber
\\ & \ \ \ \ \ \ \ \ \ \ \ \ \ \ \
\times \big(p_1^{\rho_1}p_1^{\rho_1'} 
p_2^{\rho_2} p_2^{\rho_2'} 
k_1^{\sigma_1} k_1^{\sigma_1'} 
k_2^{\sigma_2} k_2^{\sigma_2'} \big) 
\nonumber
\\ & \ \ \ \ \ \ \ \ \ \ \ \ \ \ \
\times g_{\alpha_1 \beta_1}
g_{\alpha_1' \beta_1'} 
g^{\alpha_1 \alpha_2}
g^{\beta_1 \beta_2}  
g^{\alpha_1' \alpha_2'} 
g^{\beta_1' \beta_2'}  
\\ & \ \ \ \ \ \ \ \ \ \ \ \ \ \ \ 
 \times
 \big(
\epsilon_{\alpha_2 \mu \rho_1 \sigma_1} 
g^{\mu\mu'}
\epsilon_{\alpha_2' \mu' \rho_1' \sigma_1'} 
\big)
\big(
\epsilon_{\beta_2 \nu \rho_2 \sigma_2} 
g^{\nu \nu'}
\epsilon_{\beta_2' \nu' \rho_2' \sigma_2'} 
\big). \nonumber
\end{align}
Now taking the photons to be on shell, any $g_{\rho_i \rho_i'}$ will end up being $p_i^2=0$ so we can drop these terms. Contracting metric tensors and simplifying 
\begin{align}
|\mathcal{\overline{M}}|^2 & = 
\frac{|\lambda_{h\gamma_D\gamma_D}|^2 |G_{a\gamma\gamma_D}|^4 }{64 
\left( (k_1^2 - m_{\gamma_D}^2)^2 + m^2_{\gamma_D} \Gamma^2_{\rm total}(\gamma_D)  \right) 
\left( (k_2^2 - m_{\gamma_D}^2)^2 + m^2_{\gamma_D} \Gamma^2_{\rm total}(\gamma_D)  \right)} 
\nonumber
\\ & \ \ \ \ \ \ \ \ \ \ \ \ \ \
\times \big(p_1^{\rho_1}p_1^{\rho_1'} 
p_2^{\rho_2} p_2^{\rho_2'} 
k_1^{\sigma_1} k_1^{\sigma_1'} 
k_2^{\sigma_2} k_2^{\sigma_2'} \big) 
\nonumber
\\ & \ \ \ \ \ \ \ \ \ \ \ \ \ \
\times
 \Big(
g_{\rho_2 \sigma_2'} g_{\sigma_2 \sigma_1} g_{\rho_1' \rho_2'} g_{\sigma_1' \rho_1} 
- g_{\rho_2 \sigma_1} g_{\sigma_2 \sigma_2'} g_{\rho_1' \rho_2'} g_{\sigma_1' \rho_1} 
+ g_{\rho_2 \sigma_1} g_{\sigma_2 \rho_2'} g_{\rho_1' \sigma_2'} g_{\sigma_1' \rho_1} 
\\ & \ \ \ \ \ \ \ \ \ \ \ \ \ \ \ \ \ \ \ \
- g_{\rho_2 \sigma_2'} g_{\sigma_2 \rho_1} g_{\rho_1' \rho_2'} g_{\sigma_1' \sigma_1} 
+ g_{\rho_2 \rho_1} g_{\sigma_2 \sigma_2'} g_{\rho_1' \rho_2'} g_{\sigma_1' \sigma_1}
- g_{\rho_2 \rho_1} g_{\sigma_2 \rho_2'} g_{\rho_1' \sigma_2'} g_{\sigma_1' \sigma_1}
\nonumber
\\ & \ \ \ \ \ \ \ \ \ \ \ \ \ \ \ \ \ \ \ \
- g_{\rho_2 \rho_1} g_{\sigma_2 \sigma_2'} g_{\rho_1' \sigma_1} g_{\sigma_1' \rho_2'}
+ g_{\rho_2 \rho_1} g_{\sigma_2 \rho_2'} g_{\rho_1' \sigma_1} g_{\sigma_1' \sigma_2'}
+ g_{\rho_2 \sigma_2'} g_{\sigma_2 \rho_1} g_{\rho_1' \sigma_1} g_{\sigma_1' \rho_2'}
\Big) .
\nonumber 
\end{align}
Dotting the momenta and simplifying
\begin{align}
|\mathcal{\overline{M}}|^2 & = 
\frac{|\lambda_{h\gamma_D\gamma_D}|^2 |G_{a\gamma\gamma_D}|^4 f(p_1, k_1, p_2, k_2)}{32 
\left( (k_1^2 - m_{\gamma_D}^2)^2 + m^2_{\gamma_D} \Gamma^2_{\rm total}(\gamma_D)  \right) 
\left( (k_2^2 - m_{\gamma_D}^2)^2 + m^2_{\gamma_D} \Gamma^2_{\rm total}(\gamma_D)  \right)} 
\end{align}
with $f(p_1, k_1, p_2, k_2)$ defined in Eq.~\eqref{eq:f}.

\subsection{Phase Space Reduction}
We can reduce the four-body phase space $d_4$ into a series of two-body phase spaces as 
\begin{align}
d_4[h (\hat{s}) & \rightarrow
\gamma(p_1) ~ a(q_1) ~ \gamma(p_2) a(q_2) ]
\\
&= d_2[h(\hat{s}) \rightarrow \gamma_D(k_1) ~ \gamma_D(k_2)] 
\frac{dk_1^2}{2\pi}
\frac{dk_2^2}{2\pi}
\nonumber
\\ & \ \ \ \ \ 
\times 
d_2[\gamma_D(k_1) \rightarrow \gamma(p_1) ~ a(q_1)]  
d_2[\gamma_D(k_2) \rightarrow \gamma(p_2) ~ a(q_2)]
\nonumber
\end{align}
where $q_1 =p_3 + p_4$ and $q_2 = p_5 + p_6$.
\subsection{Narrow Width Approximation}
Now if we take the NWA for the dark photon propagators we get
\begin{align}
| \mathcal{\overline{M} }|^2 & \approx 
\frac{ \pi^2 |\lambda_{h\gamma_D\gamma_D}|^2 |G_{a\gamma\gamma_D}|^4  }
{32 m_{\gamma_D}^2 \Gamma_{\text{total}}^2(\gamma_D) } \delta(k_1^2 - m_{\gamma_D}^2) \delta(k_2^2 - m_{\gamma_D}^2) 
f(p_1, k_1, p_2, k_2) .
\end{align}
Then we can use the $\delta$ functions to get rid of the integrals over $k_1$, $k_2$ in the Lorentz invariant phase space. 
\begin{align}
d_4
& \approx \frac{1}{4\pi^2}
d_2[h(\hat{s}) \rightarrow \gamma_D(k_1) ~ \gamma_D(k_2)] 
\nonumber
\\ & \ \ \ \ \ 
\times 
d_2[\gamma_D(k_1) \rightarrow \gamma(p_1) ~ a(q_1)] 
d_2[\gamma_D(k_2) \rightarrow \gamma(p_2) ~ a(q_2)] 
\nonumber
\\ & \approx
\frac{1}{4 \pi^2 (4\pi)^{6} }
\lambda^{\frac{1}{2}}\Big(1, \frac{m_{\gamma_D}^2}{m_h^2}, \frac{m_{\gamma_D}^2}{m_h^2}\Big) 
\lambda\Big(1, \frac{m_a^2}{m_{\gamma_D}^2}, 0\Big) 
d\Omega_h d\Omega_{\gamma_D(k_1)} d\Omega_{\gamma_D(k_2)}
\end{align}
where the solid angles correspond to the individual two-body decays and all momenta are on shell. Here, $\lambda(x,y,z)$ is the two-body kinematic function given by $\lambda(x,y,z) = (x - y - z)^2 - 4yz$. 

\subsection{Removing Couplings}
Let us remind ourselves of the on-shell branching ratios for $h \rightarrow \gamma_D \gamma_D$ and $\gamma_D \rightarrow \gamma a$.
\begin{align}
\text{BR} ( h\rightarrow \gamma_D \gamma_D) & =
\frac{1}{128 \pi m_h \Gamma_{\text{total}}(h)}
|\lambda_{h\gamma_D\gamma_D}|^2 
\Big( 12 - 4 \frac{m_h^2}{m_{\gamma_D}^2} + \frac{m_h^4}{m_{\gamma_D}^4}\Big)
\lambda^{\frac{1}{2}}\Big(1, \frac{m_{\gamma_D}^2}{m_h^2},\frac{m_{\gamma_D}^2}{m_h^2}\Big)
\\
\text{BR} (\gamma_D \rightarrow \gamma a)  &= 
\frac{1}{96 \pi m_{\gamma_D} \Gamma_{\text{total}}(\gamma_D)} |G_{a\gamma \gamma_D}|^2 (m_{\gamma_D}^2 - m_a^2)^2
\lambda^{\frac{1}{2}}\Big(1, \frac{m_a^2}{m_{\gamma_D}^2}, 0\Big)
\end{align}

We can then write the branching ratio as 
\begin{align}
\text{BR}\big( &h \rightarrow \gamma_D \gamma_D \rightarrow ( a \gamma )( a \gamma ) \big)
=  \frac{\int d\Gamma}{\Gamma_{\text{total}}(h) } \nonumber
\\ &=
 \frac{ |\lambda_{h\gamma_D\gamma_D}|^2 |G_{a\gamma\gamma_D}|^4 }
{2^{20} \pi^6 m_h m_{\gamma_D}^2 \Gamma_{\text{total}}(h)  \Gamma_{\text{total}}^2(\gamma_D)}
\lambda\Big(1, \frac{m_a^2}{m_{\gamma_D}^2}, 0\Big) \lambda^{\frac{1}{2}}\Big(1, \frac{m_{\gamma_D}^2}{m_h^2},\frac{m_{\gamma_D}^2}{m_h^2}\Big)
\nonumber
\\ & \ \ \ \ \ \ \ \ \ \ \ \ \ \ 
\times 
\int f(p_1, k_1, p_2, k_2)d\Omega_h d\Omega_{\gamma_D(k_1)} d\Omega_{\gamma_D(k_2)}
\\ &=
\text{BR}(h\rightarrow \gamma_D\gamma_D)~\text{BR}(\gamma_D\rightarrow a \gamma)^2 \frac{9 ~ \int
f(p_1, k_1, p_2, k_2)d\Omega_h d\Omega_{\gamma_D(k_1)} d\Omega_{\gamma_D(k_2)}}
{(2 \pi)^3 \Big(12 - 4 \frac{m_h^2}{m_{\gamma_D}^2} + \frac{m_h^4}{m_{\gamma_D}^4}\Big)  (m_{\gamma_D}^2 - m_a^2)^4 } .
\nonumber
\end{align}

\bibliographystyle{utphys}
\bibliography{draft.bib}

\end{document}